\documentclass[12pt,a4paper,twoside]{report} 
\usepackage[utf8]{inputenc}
\usepackage[vmargin=20mm,hmargin=25mm]{geometry}
\usepackage[backend=biber,giveninits=true,bibstyle=numeric]{biblatex}
\usepackage[toc]{appendix}
\usepackage{titlesec}
\usepackage{listings}
\usepackage{graphicx}
\usepackage{fancyhdr}
\usepackage[destlabel]{hyperref}
\usepackage{anyfontsize}
\usepackage{setspace}
\usepackage{placeins}
\usepackage{xcolor}
\usepackage[inline,shortlabels]{enumitem}
\usepackage[most]{tcolorbox}
\usepackage{subfig}
\usepackage[justification=centering]{caption}
\usepackage{bold-extra} 
\usepackage{textcomp}
\usepackage{gensymb}
\usepackage{pdfpages}
\usepackage[all]{hypcap} 

\newif\ifsubmission

\newcommand{\todo}{\null}
\graphicspath{ {images/} }
\title{Spatial Name System}
\newcommand{\subtitle}{An argument for rethinking naming in the Internet\\ to enable embodied virtuality}
\author{Ryan Gibb}
\newcommand{\supervisor}{Anil Madhavapeddy}
\newcommand{\cosupervisor}{Jon Crowcroft}
\date{June 6, 2022}


\titleformat{\chapter}[display]{}{\large\textsc{CHAPTER} \thechapter}{3em}{\hrule\vspace{0.5ex}\huge}[\hrule]

\hypersetup{
	colorlinks=true,
	linkcolor=link,
	anchorcolor=black,
	citecolor=black,
	menucolor=black,
	filecolor=magenta,
	urlcolor=link,
}

\emergencystretch=\maxdimen
\setlist{nosep}

\AtEveryBibitem{
	\clearfield{urlyear}
	\clearfield{urlmonth}
}

\definecolor{link}{HTML}{0066cc}

\definecolor{code-bg}{HTML}{f2f2f2}
\definecolor{code-lines}{HTML}{d8d8d8}
\definecolor{code-frame}{HTML}{bfbfbf}

\newtcblisting{framedlisting}[3][]{
	listing only,
	enhanced,
	overlay broken={
		top=0mm,
		bottom=0mm,
	},
	sharp corners,
	colback=code-bg,
	colframe=code-frame,
	boxrule=0.5pt,
	left skip=\parindent,
	right skip=\parindent,
	before skip=4mm,
	after skip=4mm,
	boxsep=0mm,
	left=1mm,
	right=1mm,
	top=0mm,
	bottom=0mm,
	title=#3,
	attach boxed title to top right={yshift=-\tcboxedtitleheight},
	boxed title style={
		sharp corners,
		rounded corners=southwest,
		colback=code-frame,
		colframe=code-frame,
		top=0mm,
		bottom=0mm,
		left=0mm,
		right=0mm
	},
	fonttitle=\color{black}\fontsize{12}{10}\scshape\selectfont,
	listing options={
		aboveskip=0pt,
		belowskip=0pt,
		basicstyle=\fontsize{10}{11}\ttfamily\selectfont,
		keywordstyle=\bfseries,
		showstringspaces=false,
		commentstyle=,
		breaklines=true,
		tabsize=4,
		#2
	},
	#1
}

\begin{document}

\onehalfspacing

\ifsubmission
	\include{sections/0.1_title_page_anonymised}
\else
	\begin{titlepage}
	\hspace*{-14mm}\includegraphics[width=65mm]{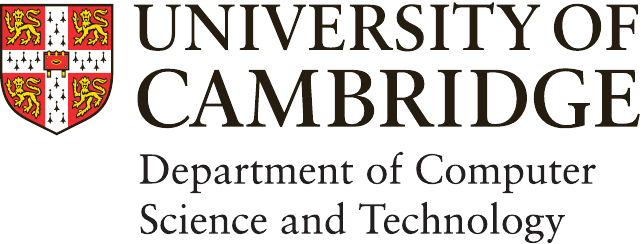}
	\begin{center}
		\makeatletter
		\vspace*{2cm}
		
		\hrule
		\vspace{0.5em}
		{
			\fontsize{30}{36}\selectfont
			\textsc{\@title}
		}
		\vspace{0.75em}
		\hrule
		
		\vspace{0.5cm}
			\large
			\textit{\subtitle}
		\vspace{1.5cm}
		
		\begin{tabular}{rl}
			\textit{Author:} & \@author\\
			\textit{Supervisor:} & \supervisor\\
			\textit{Cosupervisor:} & \cosupervisor \\
		\end{tabular}
		\vfill
		
		
		\includegraphics[width=0.35\textwidth]{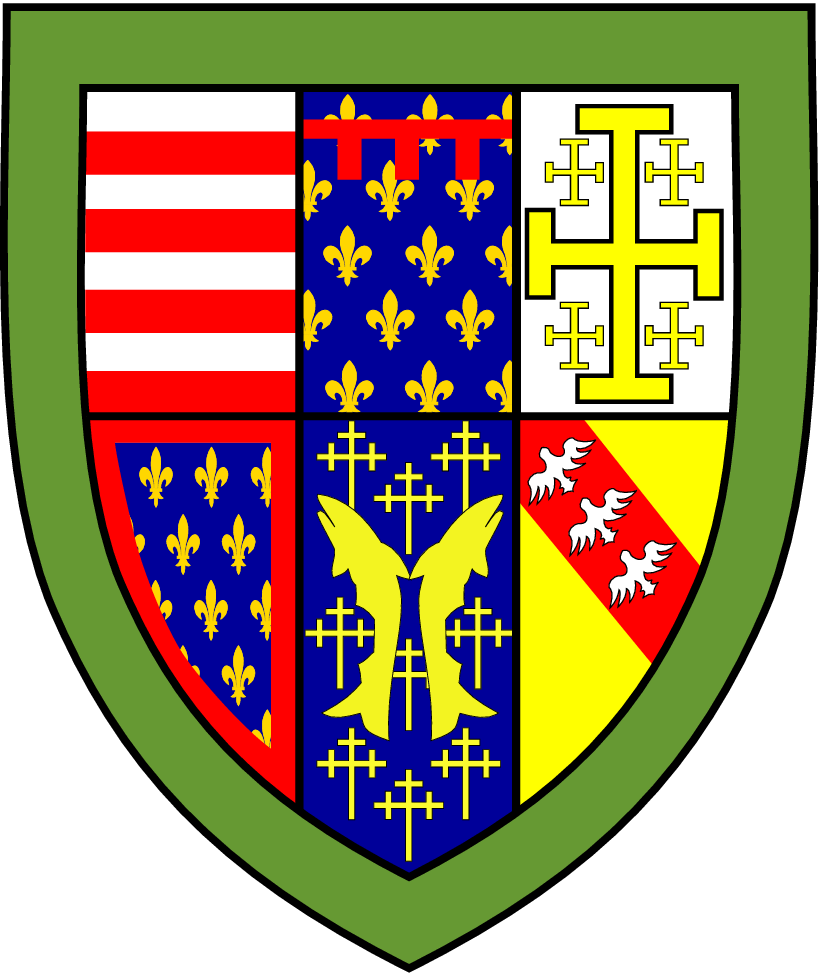}
		
		\Large
		Department of Computer Science and Technology\\
		Queens' College\\
		University of Cambridge\\
		\@date
		
		\makeatother
	\end{center}
\end{titlepage}
\fi

\newcommand{\texcount}[1]{
	\immediate\write18{texcount -merge -sum -q -1 #1.tex > #1.wc }
	\input{#1.wc}
}

\newcommand{\detailtexcount}[1]{
	\immediate\write18{texcount -merge -sum -q #1.tex > #1.wcdetail }
}

\thispagestyle{plain}

\makeatletter
\@tempcnta=\getpagerefnumber{lastpage}\relax%
\advance\@tempcnta by -\getpagerefnumber{ch-introduction}%
\advance\@tempcnta by 1%
\xdef\contentpages{\the\@tempcnta}%
\makeatother

\noindent
Main chapters:
\contentpages~pages
(pp~\pageref{ch-introduction}--\pageref{lastpage})

\noindent
Main chapters word count:
	\immediate\write18{texcount -merge -sum -q -1 main.tex > main.wc }
	\input{main.wc}

\detailtexcount{main}

\noindent
Methodology used to generate that word count:
\begin{lstlisting}
	$ texcount -merge -sum -q -1 main.tex
\end{lstlisting}

\ifsubmission\else
	\chapter*{Declaration}
\label{ch-declaration}

I, Ryan Gibb of Queens' College, being a candidate for the MPhil in Advanced Computer Science hereby
declare that this report and the work described in it are my own work,
unaided except as may be specified below, and that the report does not
contain material that has already been used to any substantial extent
for a comparable purpose.


\bigskip 
\textbf{Signed: Ryan Gibb}

\bigskip
\textbf{Date: June 6, 2022}
\vspace{\fill}

\fi

\chapter*{Abstract}
\label{ch-abstract}

The development of emerging classes of hardware such as Internet of Thing devices and Augmented Reality headsets has outpaced the development of Internet infrastructure.
We identify problems with latency, security and privacy in the
	global hierarchical distributed Domain Name System.
To remedy this, we propose the Spatial Name System, an alternative network architecture that relies on the innate physicality of this paradigm.
Utilizing a device's pre-existing unique identifier, its location, allows us to identify devices locally based on their physical presence.
A naming system tailored to the physical world for ubiquitous computing can enable reliable, low latency, secure and private communication.


\ifsubmission\else
	\chapter*{Acknowledgements}
\label{ch-acks}

I would like to thank Anil for his invaluable advice and enthusiasm during this project; Jon for inspiring me to work on a project that rethinks the architecture of our Internet; my brother for always lending an ear; and my mother, father, and grandma for encouraging me to pursue higher education.

\fi

\hypersetup{linkcolor=black}
\tableofcontents
\hypersetup{linkcolor=link}

\chapter{Introduction}
\label{ch-introduction}




We're a far cry from the sleek frictionless vision of computing science fiction has promised us for decades.
In this world, there are no phones, no laptops and no dedicated personal computing devices.
Instead, our environment is imbued with computers that blend into the background and become `invisible'.
One can interact with tables and walls to perform common tasks that we would currently use a laptop or phone for.
Already we're seeing `Internet of Thing' devices permeating our environment.
However, they have been uninspired compared to this vision, in large part limited to Internet-connected sensors and appliances, such as smart speakers, smart fridges, etc.
They don't combine to create the greater whole we envisioned and are plagued by usability, reliability, security and privacy issues.
Mark Weiser talked about this vision, coined `ubiquitous computing' in the late 1980s~\cite{weiserComputerScienceIssues1993}.
30 years later it is yet to be realised, but the hardware and software required to support this vision have made great strides in recent decades, such as small energy-efficient embedded computers, machine learning models for tasks gesture recognition, and augmented reality headsets.


Mark Weiser also referred to ubiquitous computing (ubicomp) as `embodied virtuality', as he considered virtual reality as diametrically opposed to his vision.
Augmented reality, however, is aligned with the principles of ubicomp; rather than aiming to create a replacement for the real world, instead augments it with computing.
Consider, if you could intuitively interact with these `invisible' computing devices in your physical environment through an AR interface.
If you could dim the lights at a glance, monitor and control what devices are recording information, and interact with the virtual world embedded in our physical world directly.

In this dissertation, we argue the thesis that:\vspace{-1em}
\begin{quotation}
	\itshape
	We have the hardware and software to support low latency augmented reality interactions, but the current network architecture is inadequate to support interconnecting them.
	We need a Spatial Name System that can map physical device locations to network addresses to overcome this limitation and unlock the potential of augmented reality.




	%
\end{quotation}

\clearpage


The Internet and its associated systems have been developed in the IETF with the principles of `rough consensus and working code'.
This has been extremely successful in getting working systems that people can benefit from in a timely manner, but it also means that systems aren't designed with a view to the long term.
Additionally, due to protocol ossification --- the difficulties in changing and modifying established standards --- the Internet is slow to evolve.
Combined, these factors mean that systems are used in ways they aren't designed or suitable for.

Looking at one particular component of the Internet architecture, the Domain Name System (DNS), can illustrate our point.
The DNS is a global hierarchical naming system used to name devices in the Internet by translating human-readable domain names to IP addresses used for identification and routing.
While this is effective in resolving names for remote devices across the Internet with administration delegated to various organisations, it is a poor fit for naming proximate physical devices.
It adds latency to requests, is slow to update, and has security and privacy implications.



This project will consider the systems support required for the use case of an AR interface into ubicomp, namely meeting the low latency requirements for naming physically proximate machines in the Internet, as well as a way to name devices based on their location.
The main artefact of this dissertation is the justification for, and the design of, an alternative naming mechanism to the DNS that is native to ubicomp, the Spatial Name System (SNS).
This protocol uses a unique identifier ubicomp devices already have, their physical location.
It acknowledges the physicality of devices and aims to provide low latency resolution to network addresses, along with considering privacy and security concerns that weren't present in the original DNS design.



We continue this dissertation in Chapter~\ref{ch-background} with necessary background material to justify {\itshape `We have the hardware and software to support low latency augmented reality interactions, but the current network architecture is inadequate to support interconnecting them.'} as well as the technologies that will enable our Spatial Name System.
In Chapter~\ref{ch-augmented-reality-interface} we will examine the work done towards creating an AR interface into the world of ubiquitous computing to justify to the claim {\itshape `We need a Spatial Name System that can map physical devices to network addresses to overcome this limitation and unlock the potential of augmented reality.'}
The deficiencies found in the existing network architecture will be explored in chapter~\ref{ch-spatial-networking}, and an alternative proposed.
Chapter~\ref{ch-related-work} will relate this design to existing work, and we will conclude with chapter~\ref{ch-conclusions}.

\chapter{Background}
\label{ch-background}

We continue this dissertation in Chapter~\ref{ch-background} with background material necessary to justify the claims that {\itshape `We have the hardware and software to support low latency augmented reality interactions'}~~(\S\ref{sec-ubicomp}), and {\itshape `but the current network architecture is inadequate to support interconnecting them'}~~(\S\ref{sec-internet-arch}), then conclude with an overview of relevant technologies that will enable our Spatial Name System (SNS)~~(\S\ref{sec-local-positioning-systems}).

\section{Ubiquitous Computing}
\label{sec-ubicomp}

\begin{figure}[t]
	\centering
	\includegraphics[width=0.75\textwidth]{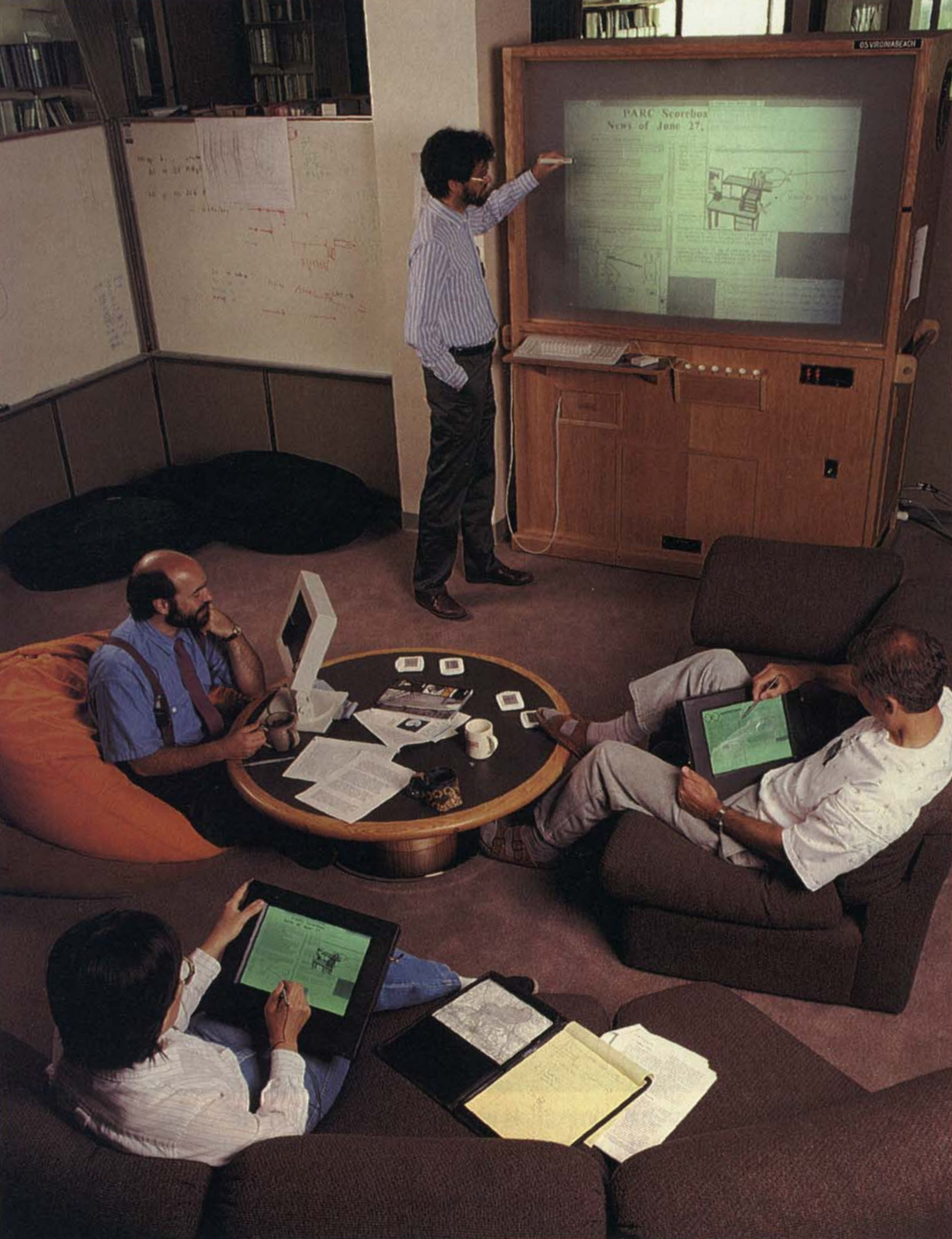}
	\caption[]{Weiser's ubicomp demonstration in 1999~\cite{weiserComputer21stCentury1999}.}
	\label{fig-ubicomp}
\end{figure}

Weiser's compelling vision of ubiquitous computing (ubicomp) envisioned a world where computers are available throughout a physical environment, but they blend into the background and are effectively invisible to the user~\cite{weiserComputerScienceIssues1993}.
He compared this to the technology of writing which permeates our environment in industrialized countries --- such as in signs, billboards, packaging and labelling --- and conveys information at a glance without requiring constant attention.
Ubicomp devices would augment our environment and allow us to focus on other things, instead of demanding even more attention like our current mobile personal devices like smartphones and smartwatches~\cite{weiserDesigningCalmTechnology1996}.
Figure~\ref{fig-ubicomp} shows a picture of Weiser and colleagues demonstrating their vision of ubicomp with the technology available at the time.

\subsection{The Internet of Things}
\label{sec-iot}

The Internet of Things (IoT) is a modern interpretation of ubicomp that imagines a vast network of `smart' devices all connected via the Internet such as appliances like light switches, thermostats, stereos, ovens, fridges and smart plugs.
Some commercial examples are the Alexa Smart Speaker, Google Home suite and Ring security system.
These devices all have something in common: they primarily rely on Internet communication.
Google Homes cannot function without connectivity to remote datacentres; there are webpages listing hundreds of security cameras hacked via their remote access functionality; and generally IoT devices, as their name suggests, are reliant on the Internet.
 
In chapter~\ref{ch-introduction} we claimed that IoT devices are `plagued by usability, reliability, security and privacy issues', which is due to their reliance on Internet infrastructure.
Regarding usability, often the interface to these devices is through an application on a smartphone, which is almost the antithesis of Weiser's vision of making computers invisible.
To address reliability, communication over the Internet is inherently unreliable due to its best effort delivery method.
Congestion control adds reliability on top of this unreliable substrate, but connectivity issues often still occur due to congestion issues, remote data centre failures and configuration changes like Meta's (formerly Facebook) BGP/DNS failure last year.
And finally, security and privacy are of utmost importance with ubicomp devices due to their pervasive nature.
Communicating sensitive data to remote data centres relinquishes privacy to the provider, but also adds another attack vector for malicious actors.
Just connecting devices to the Internet has serious implications for privacy and security~\cite{UninvitedInternetThings2021}.

Part of the reason for these problems is the architecture associated with IoT devices, like relying on remote data centre processing.
But digging deeper we argue the reason for these architectures is because of the inadequacy of the existing infrastructure and the lack of a viable alternative.
This has resulted in a divergence from Weiser's original vision which focused on user interaction and local communication.
Why should ubicomp require Internet connectivity in order to do something that by all accounts only needs to work locally when it has so many negative effects?
Instead, we can remove the Intenet-centric design of IoT devices in order to better realise the original vision of ubicomp~\cite{madhavapeddyArchitectureInterspatialCommunication2018}.



Considering the drawbacks of the Internet of Things, in this dissertation, we shall consider how we can better design alternative systems to better support ubicomp.
For devices that transmit and display information directly, like devices with screens, Weiser identified two issues of crucial importance: location and scale.
He said that ubiquitous computers must know where and what they are in order to adapt their behaviour in significant ways without requiring any more sophisticated techniques like artificial intelligence~\cite{weiserComputer21stCentury1999}. 
We would extend this argument to say that location and scale are also important for ubicomp devices in order for people to be able to interact with them directly without awkward in-betweens like using an app on a smartphone.
We will explore the role location can play not just in informing devices of their environment, but also in how to identify devices based on their location with the SNS in chapter~\ref{ch-spatial-networking}.

\subsection{Augmented Reality}
\label{sec-augmented-reality}


While the hardware developments we have been covering so far have been devices directly relating to ubicomp, other developments have taken place in virtual and augmented reality.
Zuckerberg has staked Meta's future on his vision of the Metaverse, a term coined by Stephenson in 1992~\cite{stephensonSnowCrash1992}, with a heavy reliance on virtual reality (VR) with their Quest headset series.
Weiser considered virtual reality to be `diametrically opposed to [their] vision' as it `attempts to make a world inside the computer'.
He said:
\begin{quotation}
	\itshape
	Indeed, the opposition between the notion of virtual reality and ubiquitous, invisible computing is so strong that some of us use the term `embodied virtuality' to refer to the process of drawing computers out of their electronic shells.
	\hfill \normalfont Weiser, 1999~\cite{weiserComputer21stCentury1999}
\end{quotation}
However, augmented reality (AR) provides an opportunity to realise this vision of embodied virtuality.
This is fundamentally different from VR which tries to replace the real world with a virtual one.
Augmented reality could provide an interface into the world of ubicomp --- bringing the invisible computers into the visible for the purposes of interaction~\cite{rehmanInterfacingInvisibleComputer2002}.
Imagine if you could control devices with gazes and gestures, instead of clunky and inconvient touchscreen displays.
If you could monitor the digital infrastructure in your environment - see what is recording your audio, what the heating policy is, and which physical devices have hidden computers inside them.
If you could obtain telemetrics for devices 
This could be for the purposes of control, monitoring telemetrics, or just for passively gathering information.

\begin{figure}[t]
	\centering
	\includegraphics[width=0.8\textwidth]{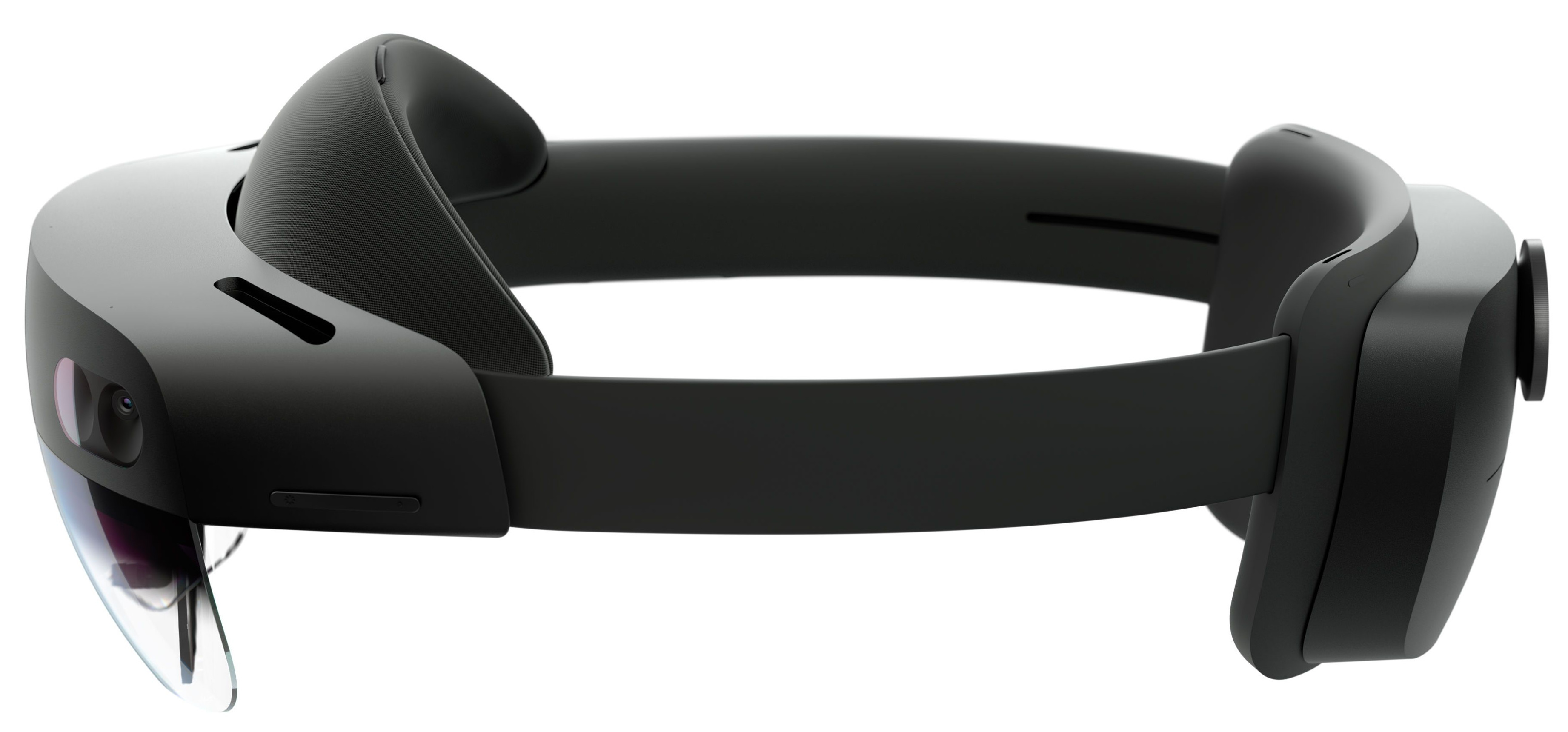}
	\caption[]{Microsoft's augmented reality headset the HoloLens 2~\cite{HoloLens}.}
	\label{fig-hololens2}
\end{figure}

Microsoft has recently released the HoloLens 2, an AR headset that projects virtual 'holograms' in front of the user's vision.
AR hardware is still in its infancy; as can be seen in figure~\ref{fig-hololens2} the headset is rather bulky and it has a limited FOV of 52 degrees.
Microsoft is mainly targetting industrial use cases with this headset, but there is renewed interest in AR across the industry, with Apple and Meta purportedly developing their own hardware.
Perhaps this hardware will develop into something unobtrusive, similar to a normal pair of eye-correcting glasses. 

One requirement of AR compared to traditional user interfaces is the demand for low latency interaction~\cite{deberHowMuchFaster2015}.
In order for interactions to feel natural ubicomp interaction requires low latency in the order of milliseconds, rather than delaying for seconds that we're used to with modern Internet applications on mobile devices.
Communication over the Internet adds latency to the operation of ubicomp devices.
For example, DNS queries may result in multiple iterative requests being made.


An alternative mode of identifying devices visually to the SNS would be to use visual tags like quick response (QR) codes~\cite{scottUsingVisualTags2005}.
But this has a few disadvantages: devices need to be large enough to fit a QR code; lighting conditions can affect resolution; they are static once created; and they are aesthetically unappealing, which may be permissible in an industrial environment but is not desirable in a home environment.

In summary, we have the hardware and software, in the form of augmented reality headsets and ubicomp devices, to support low latency augmented reality interactions.
However, to network these devices, we argue the current Internet architecture is inadequate.

\clearpage
\section{Internet Architecture}
\label{sec-internet-arch}

We have claimed that the current network infrastructure is inadequate to support our vision of an AR interface in the world of ubicomp.
In this section, we will justify this claim by considering existing protocols that would be candidates for networking our AR interface.

Consider the simplest possible scenario for this interface: connecting to a local smart device from an AR headset.
In this example, we only have two devices: one (the headset) trying to resolve the network address of the other.
At the lowest level of the Internet protocol suite, we need a Media Access Control (MAC) address for link layer communication technologies like Ethernet, IEEE 802.11 (Wi-Fi) and Bluetooth.

\subsection{The Address Resolution Protocol}
\label{sec-arp}

The Address Resolution Protocol (ARP) resolves link layer MAC network addresses to Internet Protocol (IP) addresses.
One may consider why not implement a 'Spatial Resolution Protocol', resolving spatial coordinates to MAC addresses.
Such a system would not provide the flexibility required for ubicomp, as it would only work on a local network or ARP-bridged LANs.
While devices may be physically proximate, they may be using a variety of network connectivity such as different IEEE 802.11 networks, Bluetooth, LoRa, or Zigbee.
Our communication could be happening within a physical space using disparate network technologies.
The physical and network topologies might not necessarily overlap.
Therefore, we need a way to refer to devices higher level.





\subsection{\texttt{HOSTS.TXT}}
\label{sec-hosts.txt}

A naming system in the Internet associates names with IP addresses, like how IP addresses are associated with MAC address with ARP but at a higher layer.
The ARPANET, the precursor to the Internet, did not have a standardised naming mechanism.
Originally each node or network maintained its own mapping of names to IP addresses, with no mechanism for distribution or consistency.
RFC 606 `Host Names On-line' by Deutsch in 1973~\cite{HostNamesOnline1973} recognises this problem and proposes a centralized solution administered by the Network Information Center (NIC), specifically Elizabeth J. Feinler, transmitted over the File Transfer Protocol (FTP),
\begin{quotation}
	\itshape
	Now that we finally have an official list of host names, it seems
	about time to put an end to the absurd situation where each site
	on the network must maintain a different, generally out-of-date,
	host list for the use of its own operating system or user
	programs.
	\hfill \normalfont Deutsch, Dec 1973~\cite{HostNamesOnline1973}
\end{quotation}
However, the limitations of this solution are also acknowledged:
\begin{quotation}
	\itshape
	I realize that there is a time-honored pitfall associated with
	suggestions such as the present one: it represents a specific
	solution to a specific problem, and as such may not be compatible
	with or form a reasonable basis for more general solutions to more
	general problems.
	However, (1) this particular problem has been
	irking me and others I have spoken to for well over a year, and it
	is really absurd that it should have gone unsolved this Long; (2)
	no one seems particularly interested in solving any more general
	problem.\par
	\hfill \normalfont Deutsch, Dec 1973~\cite{HostNamesOnline1973}
\end{quotation}

The next year, RFC 608 defines the server from which this `\texttt{HOSTS.TXT}' file should be downloaded from via FTP at `OFFICE-1' (address 43) with username `GUEST' and password `ARPA'~\cite{HostNamesOnline1974}.
One can still edit this file at \texttt{/etc/hosts} in a modern UNIX system to override name resolution.
However, this was an ad hoc solution to the problem at hand and did not anticipate problems that would arise in later years, such as authentication, privacy and, for our case, latency.
Most imminently, however, was scalability as the network rapidly grew.
Standardization efforts and distribution protocols could only take this centralized system so far and eventually this necessitated the creation of the distributed Domain Name System.

\subsection{The Domain Name System}
\label{sec-dns}

\begin{figure}[b]
	\centering
	\includegraphics[width=0.5\textwidth]{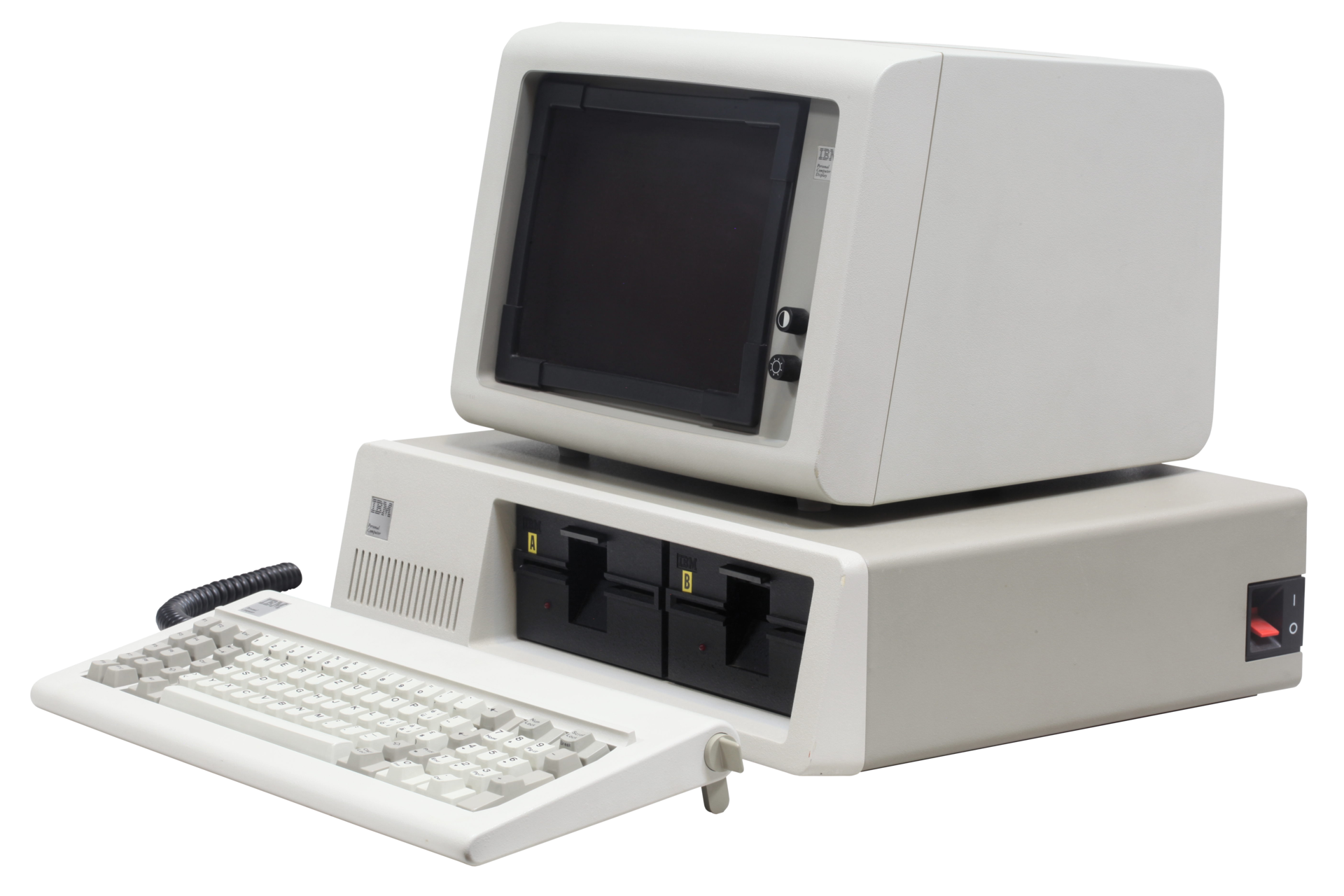}
	\caption{IBM Personal Computer model 5150~\cite{boloEnglishIBMPersonal2014}.}
	\label{fig-ibm}
\end{figure}

The Domain Name System (DNS) is the hierarchical decentralized system that is the solution to distributed naming in the Internet.
In this context, hierarchical means that a domain has subdomains that can be delegated to other authorities.
Top-Level Domains (TLDs), such as \texttt{.com}, \texttt{.org}, and \texttt{.net} are the highest level domains below the root domain.
Decentralised means that each delegated authority has an authoritative nameserver(s).
DNS resolvers follow the chain of delegation to find this.
While DNS is distributed and the administration is delegated, the system is still unified; relying on 13 root nameservers, or rather 13 root nameservers addresses, that return the authoritative nameservers for the TLDs.
In some ways, it is simply an iteration on the \texttt{hosts.txt} file to ensure scalability.

The computers for which the DNS was developed were heavy, large and static.
The IMB Personal Computer Model 5150, see figure~\ref{fig-ibm}, was released in 1982, the year before domains names were published~\cite{DomainNamesConcepts1983}, weighting over 9kg.
While DNS has been successful in addressing physically remote computers belonging to an organisation over the Internet,
it is not an appropriate naming mechanism for physical devices.

\begin{figure}[t]
	\centering
	\includegraphics[width=1\textwidth]{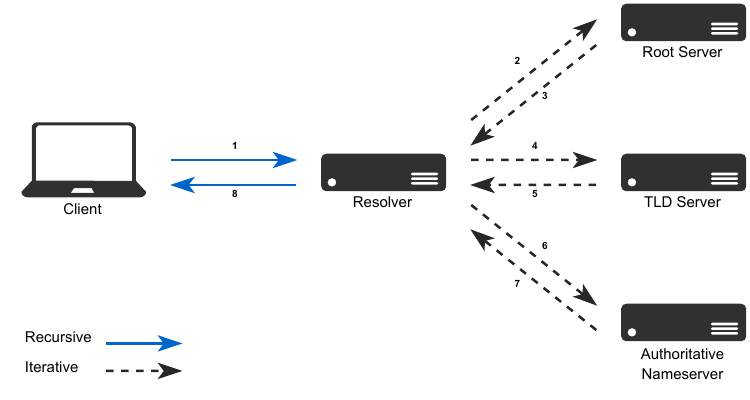}
	\caption[]{An example of the requests associated with a DNS lookup.}
	\label{fig-dns-lookup}
\end{figure}

First of all, there is a latency associated with making requests across the Internet which is generally correlated with geographical distance.
This is compounded by applications' numerous sequential requests.
See figure~\ref{fig-dns-lookup} for an example of the steps involved in a DNS lookup.
Caching is used to improve performance, but this has a tradeoff with freshness.
DNS updates take time to propagate through this distributed hierarchy.
Setting time-to-live (TTL) values is a delicate balance of request latency, server load and response freshness.
This is a problem for mobile devices that require both fresh and low latency responses.

Communicating across the Internet also has reliability implications; in many cases, a failure in DNS can appear as a failure in the network.
Take Meta's BGP/DNS failure last year that was mentioned in section~\ref{sec-ubicomp}~\cite{UnderstandingHowFacebook2021}.
Aside from the technical details (which are interesting in their own right), this resulted in the situation that employees were unable to physically access office spaces due to remote Internet architecture failures~\cite{taylorFacebookOutageWhat2021}.
This is a spectacular failure in ubicomp due to an overreliance on Internet architecture, but it is not the only issue with DNS.

As mentioned, authentication and privacy been retrospectively fitted to DNS.
DNSSEC provides a set of security extensions to DNS to authenticate responses, so a resolver knows that the response was signed by the holder of a private key.
There is a hierarchical authentication model with each zone relying on its parent zone to authenticate it. 
The root domain is signed in a root signing ceremony, where 7 people from ICANN and other trusted Internet community members sign the root zone's public keying information.
The trust in DNSSEC is derived from the security protocols put in place around this ceremony.
For privacy, DNS over TLS (DoT) / DNS over HTTPS (DoH) can be used.
However, deployment of these protocols is limitted, due to protocol ossification.
DNSSEC has a lot of overhead for our use case, and it relies on the security of a remote system that we have no control over.
Finally, DNS doesn't provide a natural way of naming ubicomp devices.
Domain names were created for expressing hierarchical organisational membership, but not for mobile local devices.





One could hack DNS for ubicomp naming, running a local authoritative nameserver, but this would be bringing a lot of operational baggage required for distribution that is not required locally as well as requiring infrastructure in the form of a server.
Instead, multicast DNS~\cite{cheshireMulticastDNS2013} (mDNS) is a solution for a decentralised (as opposed to distributed) DNS system, created to serve Zero-configuration networking (zeroconf).
\begin{quotation}
	\itshape
	As networked devices become smaller, more portable, and more\\
ubiquitous, the ability to operate with less configured
infrastructure is increasingly important.
	\hfill \normalfont RFC 6762 Multicast DNS, 2013~\cite{cheshireMulticastDNS2013}
\end{quotation}
It provides the same APIs as DNS, but with a different implementation.
Instead of querying a nameserver, all participants communicate directly using IP/UDP multicast.
By default mDNS exclusively uses domains with the \texttt{.local} TLD.
When a client queries a domain it sends a multicasted query packet to a reserved multicast address.
The response is typically multicasted too, so other participants can cache the response.
This means that its scope is limited to a multicast domain, typically a local network, and not over the Internet as a whole.
Multicast DNS is extremely effective for local service discovery but is better suited for service discovery than for low latency spatial communication.

There are a number of problems with mDNS for our use cases,
\begin{enumerate*}[(1)]
	\item there is no authentication, privacy considerations, or access control;
	\item it only works within a multicast domain; and
	\item names are descriptive strings, and name conflict resolution relies on programmatic updates, which doesn't provide a reliable or intuitive interface for ubicomp devices.
\end{enumerate*}


To summarise, we have considered the existing address resolution and naming architectures ARP, \texttt{HOSTS.TXT}, the DNS and mDNS.
However, none of these meets the needs of our AR interface due to internetworking, latency, reliability and naming requirements.
Instead, will consider how a Spatial Name System~~(\S\ref{ch-spatial-networking}) can be designed to meet these needs using devices' physical locations. 

\clearpage
\section{Local Positioning Systems}
\label{sec-local-positioning-systems}





Weiser said that `today's computers' (computers of 1999) had no idea of their location or surroundings~\cite{weiserComputer21stCentury1999}.
This is not quite true in 2022, with global navigation satellite systems (GNSS) --- like GPS or GLONASS --- capable smartphones and wearables fitted with an array of sensors including magnetometers.
Applications can also infer location from IP addresses and WiFi access points (AP) which are associated with geographical coordinates.
However, the precision and accuracy of this association are limited.
And GPS has limited accuracy indoors and in dense urban areas~\cite{tariqNonGPSPositioningSystems2017}.
Centimetre location accuracy is still lacking in most modern devices.

\begin{figure}[h!]
	\centering
	\includegraphics[width=0.7\textwidth]{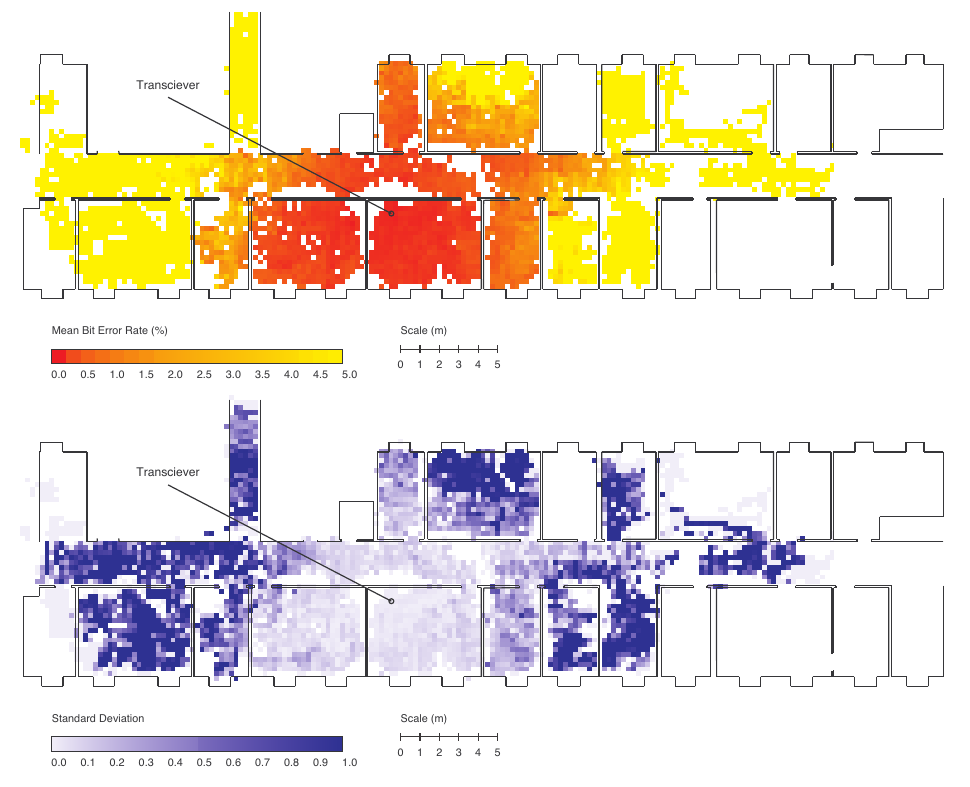}
	\caption[]{Measurement of Bluetooth signals strengths (\textit{top}) and the standard deviation (\textit{bottom}) using location from the Active BAT system in the Cambridge Computer Lab first floor north corridor~\cite{madhavapeddyStudyBluetoothPropagation2005}}
	\label{fig-cl-active-bat}
\end{figure}

There are a number of indoor positioning systems available~\cite{tariqNonGPSPositioningSystems2017, priyanthaCricketLocationsupportSystem2000}, such as the Cambridge Computer Lab Active BAT system in the early 2000s~\cite{harterAnatomyContextAwareApplication2002}.
This system provides sufficient accuracy for locating ubicomp devices; `95\% of readings are within 9cm of their true positions'~\cite{harterAnatomyContextAwareApplication2002}.
Figure~\ref{fig-cl-active-bat} shows measured Bluetooth signal strengths with varying location, where the location was obtained from the Active BAT system~\cite{madhavapeddyStudyBluetoothPropagation2005}.
The Bluetooth signal strengths are not of interest to us, but the accuracy of tracking mobile devices in the Active BAT system is demonstrated in this work.
To make devices self-aware of their location a beacon associated with one of these local positioning systems could be used.
This live location tracking would allow devices to auto-configure their location and support mobile devices~~(\S\ref{sec-mobility}).
However, these systems are very expensive to install; statically surveying immobile devices will also work.
In essence - telling devices their location instead of making them location-aware.

\clearpage
\section{Summary}
\label{sec-background-summary}

We began this chapter by discussing Weiser's vision of ubiquitous computing~(\S\ref{sec-ubicomp}), describing where we have been led astray with the Internet of Things~(\S\ref{sec-iot}), and describing how augmented reality relates to this vision as embodied virtuality~(\S\ref{sec-augmented-reality}).
We then how the current Internet infrastructure is inadequate for our vision of an augmented reality interface into the world of ubicomp~(\S\ref{sec-internet-arch}), describing address solution mechanisms~(\S\ref{sec-arp}), how the current naming system came to be~(\S\ref{sec-hosts.txt}), and why the DNS is insufficient~(\S\ref{sec-dns}).
We concluded with a summary of local positioning systems as an enabling technology for our Spatial Name System~(\S\ref{sec-local-positioning-systems}).

\newcommand{\hash}{\scalebox{0.8}{\raisebox{0.4ex}{\#}}}

\chapter{Augmented Reality Interface}
\label{ch-augmented-reality-interface}


This chapter will test the thesis that {\itshape `We have the hardware and software to support low latency augmented reality interactions'}. We will describe the technologies selected~(\S\ref{sec-hw-and-sw}), explore how to develop for augmented reality~(\S\ref{sec-ar-development}), and evaluate our experience trying to network the HoloLens for an Augmented reality interface~(\S\ref{ch-introduction}), justifying the claim {\itshape `We need a Spatial Name System that can map physical devices to network addresses to overcome this limitation and unlock the potential of augmented reality'}~~(\S\ref{sec-interactivity-evaluation}).

\section{Hardware and Software}
\label{sec-hw-and-sw}


In this section, we will describe the hardware and software artefacts chosen to advance the thesis {\itshape `We have the hardware and software to support low latency augmented reality interactions'}.

The state of the art hardware for augmented reality (AR) is the HoloLens 2~(\S\ref{sec-ubicomp}).
The HoloLens 2 was released by Microsoft in 2019 as a successor to its original HoloLens headset from 2016.
Its main focus is on the commercial market, including manufacturing, engineering, healthcare and education ---
perhaps due to its \$3,500 price tag.
To develop applications towards these ends, 2-dimensional applications are supported through the Universal Windows Platform (UWP);
UWP also supports other platforms running the Windows Operating System like desktops, laptops and tablets.
But the primary supported development environment for the headset through game engines for 3-dimensional applications including Unity and Unreal.
We choose to develop with Unity as it's been supported for the longest and has the best introductory documentation.
It is possible to program directly using DirectX and Windows APIs, but this is a significant amount of work~\cite{thetuvixNativeDevelopmentOverview}.

The development experience using these game engines is very similar to that of VR.
In fact, Microsoft describes the headset as mixed reality (MR), as opposed to virtual or augmented reality.
MR refers to a hybrid of VR and AR, a term introduced by Milgram in his reality--virtuality continuum which describes the spectrum(s) between virtual and augmented reality~\cite{milgramAugmentedRealityClass1994}.
Whereas VR is entirely virtual with no interaction with the physical world, and AR overlays the physical world with a virtual interface, MR combines both.
Our motivation for using this device is for an AR interface, however.

\begin{figure}[t]
	\centering
	\includegraphics[width=1\textwidth]{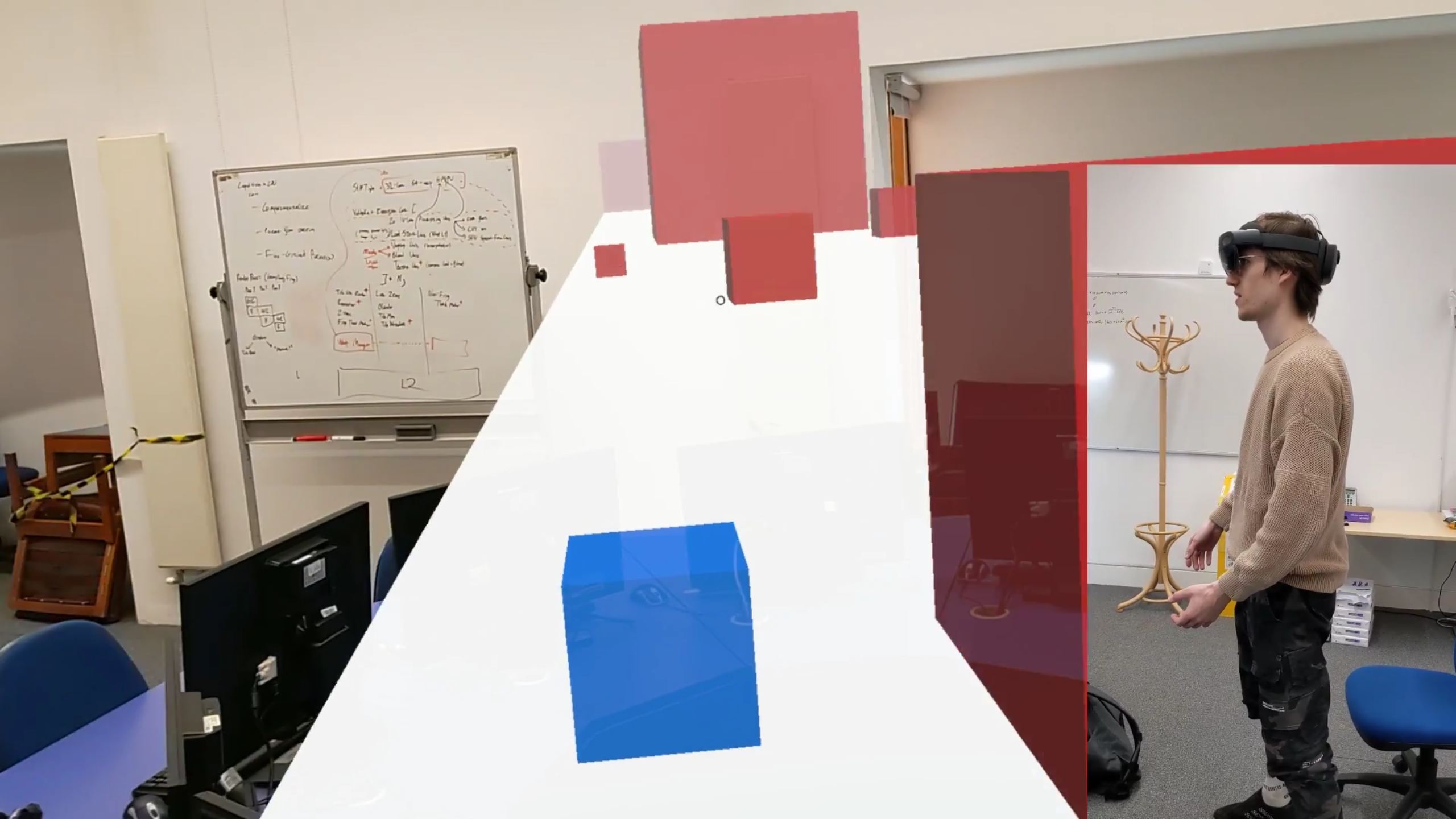}
	\caption[]{A person playing the Cubes game on the HoloLens 2.}
	\label{fig-hololens-view}
\end{figure}

In order to test the thesis {\itshape `We have the hardware and software to support low latency augmented reality interactions.'} and profile AR's latency requirements, we created an application for the HoloLens in Unity~\cite{gibbCubesGithub2022}\footnote{The game was ported VR and AR from an existing game written by the author~\cite{VirtualAugmentedReality}.}.
\todo{what are the latency requirements?}
As Unity is first and foremost a game engine, we created a game.
The purpose of this game was to explore development for the HoloLens to see if our thesis is justifiable.
This game is particularly suited to explore low latency interaction, it's an `infinite runner,' requiring the player to dodge cubes at speed.
A player's view, and a view of the player, can be seen in figure~\ref{fig-hololens-view}.
A web version is available to play~\cite{gibbCubes}.

\section{Augmented Reality Development}
\label{sec-ar-development}

In this section, we'll explore the development experience for augmented reality to test our thesis.
We will describe the development process to program a 3D project for the HoloLens~(\S\ref{sec-development-process}), the control mechanism we added to our game for MR~(\S\ref{sec-mixed-reality-control}) and attempt to hack (in the original, positive, definition of the word~\cite{HackersHeroesComputer}) the operating system of the HoloLens~(\S\ref{sec-holographic-terminal}).



\clearpage
\subsection{Development Process}
\label{sec-development-process}

\begin{figure}[b!]
	\centering
	\makebox[\textwidth]{
		\begin{tabular}{c}
					\subfloat[Unity development environment.\label{fig-unity-dev}]{\includegraphics[width = 0.7\textwidth] {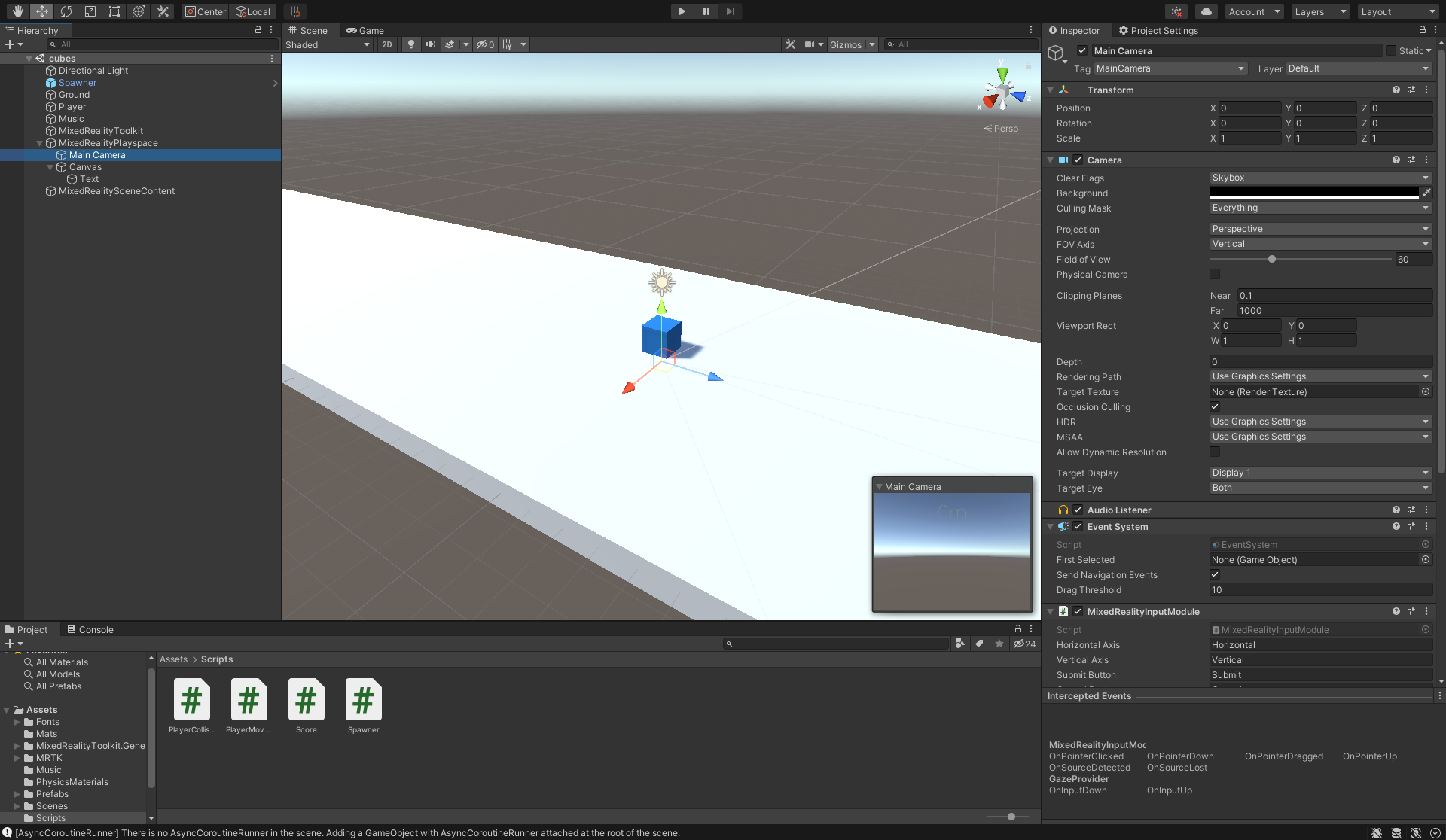}} \\
					\subfloat[Configuring remote execution in Visual Studio.\label{fig-vs-dev}]{\includegraphics[width = 0.7\textwidth] {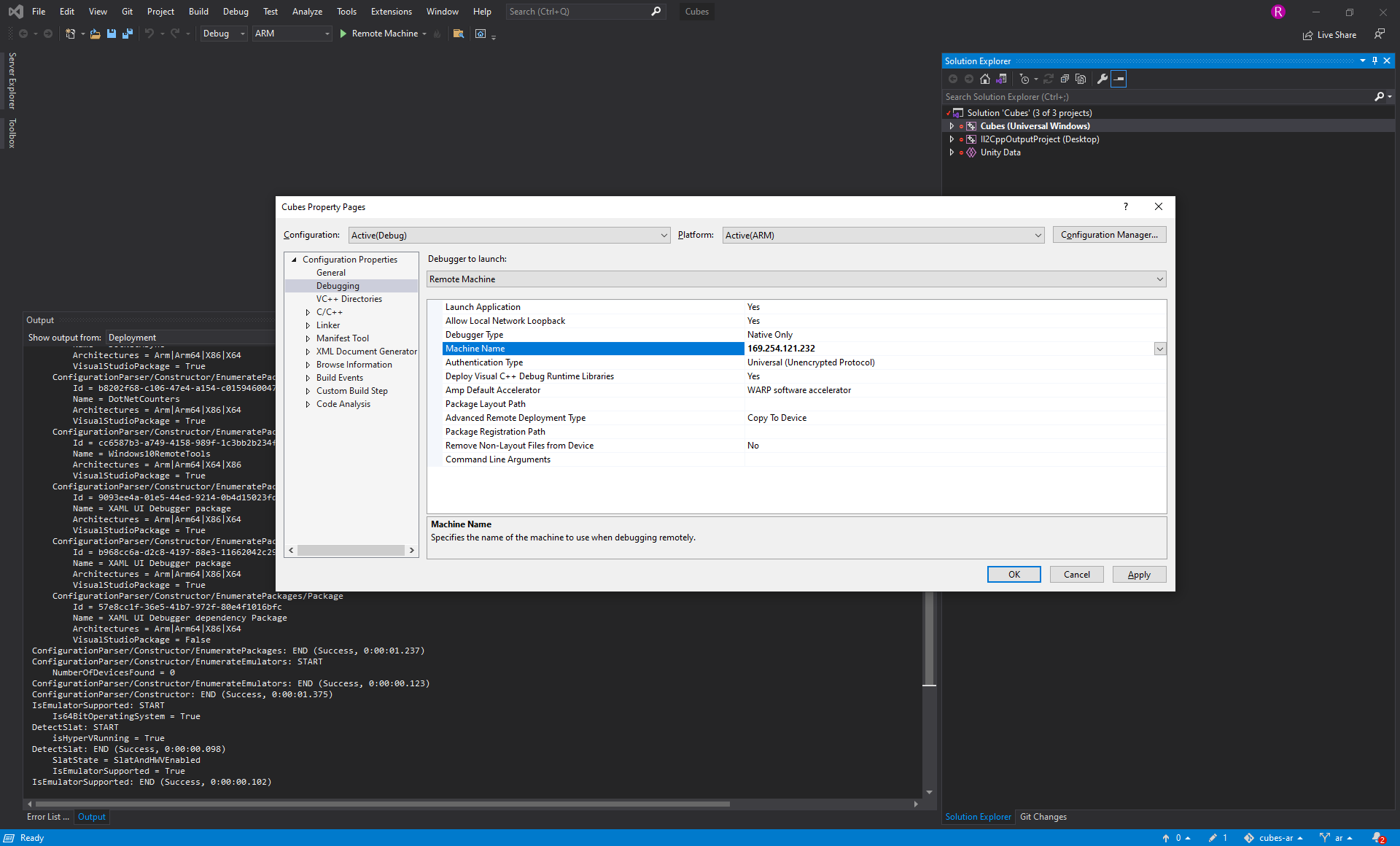}} \\
		\end{tabular}
	}
	\caption{Developing for the HoloLens in Unity (\textit{left}) and Visual Studio (\textit{right})}
	\label{fig-hololens-dev}
\end{figure}

To being, we downloading Unity Hub with Unity 2020.3.x/2019.4.x.
Unity, a game engine, was launched in 2005 to make game development more accessible.
It originally only supported Mac OS X but expanded into other platforms, and now supports Windows Mixed Reality.
Development is done through a GUI editor where the user can create and edit objects, as well as program scripts primarily in C\hash.
Unity editors are installed through the Unity Hub, as well as Unity packages.
There are additional dependencies to develop a HoloLens application, as documented on the Microsoft Mixed Reality documentation.~\cite{thetuvixInstallToolsMixed}.
This includes Windows 10 SDK 10.0.18362.0 or later, the .NET Desktop Runtime 5.0,
Visual Studio 2022 with associated `workloads', and the Microsoft Mixed Reality Toolkit (MRTK) package.
Unity additionally requires the Universal Windows Platform Build Support module and Mixed Reality Feature Tool.
Windows 10 or 11 are the only supported operating systems.

\begin{figure}[t!]
	\centering
	\includegraphics[width=1\textwidth]{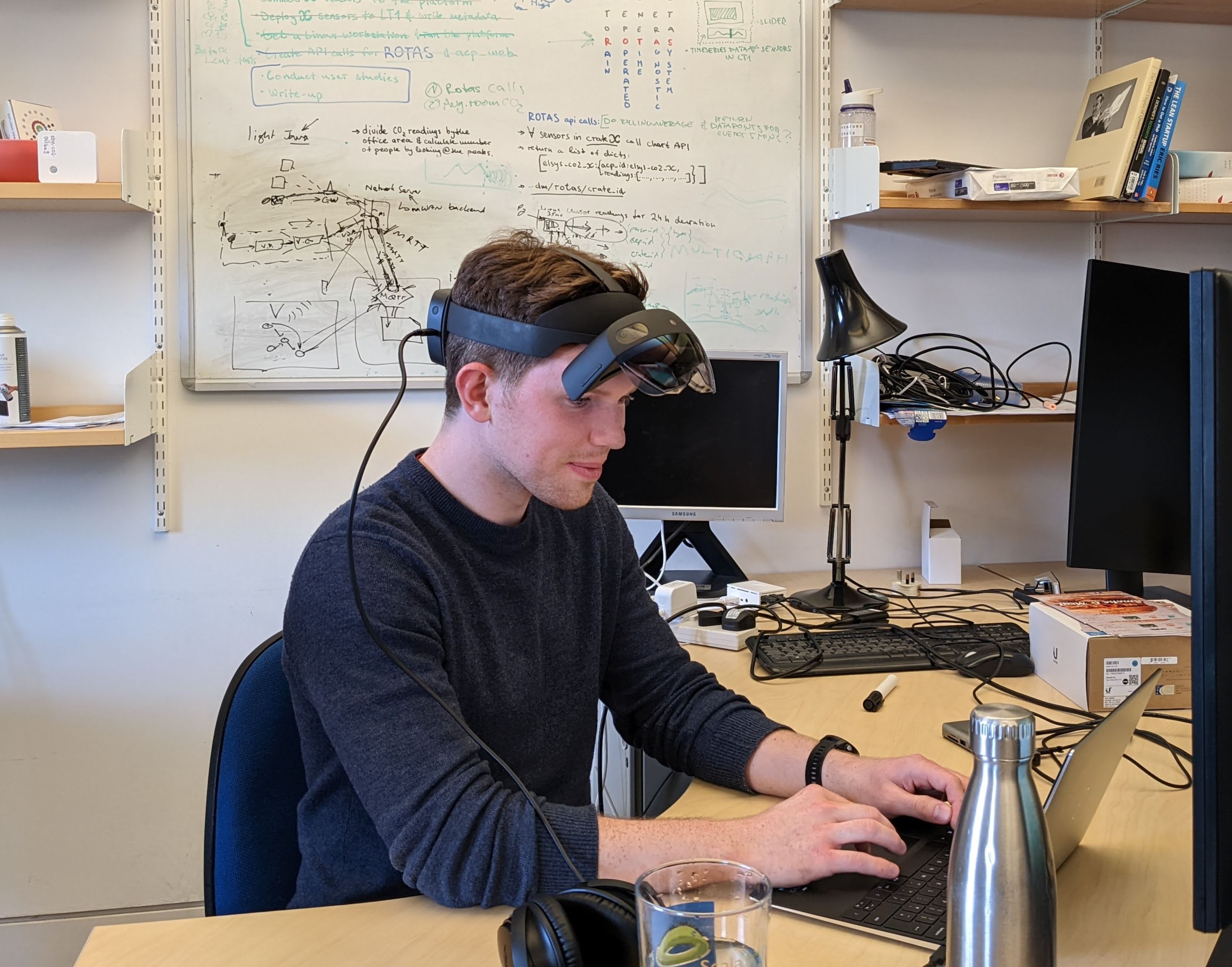}
	\caption[]{A person developing for mixed reality in mixed reality:\\wearing the HoloLens while programming it.}
	\label{fig-hololens-mr-dev}
\end{figure}

The game is developed in Unity with the MRTK package.
This development kit provides an API to use from within Unity, in the form of Unity scene objects, components and APIs.
As shown in figure~\ref{fig-unity-dev}, the Unity editor is a GUI where scenes are manipulated, objects are added to scenes, and components are added to objects.
Components add functionality to objects, like physical interactions, cameras to view the scene from, or through user-defined scripts.
More details about development with Unity can be found through the Unity documentation~\cite{UnityTutorial}, but suffice to say programming for the HoloLens is similar to programming any other game, but with additional APIs models and interaction models.
While MRTK can target Windows Mixed Reality, OpenXR is a more portable and open platform.

OpenXR is an open and royalty-free standard for virtual and augmented introduced in 2019 by the Khronos Group consortium (including Google, Apple, ARM, Intel, Qualcomm, AMD, NVIDIA, Huawei, Epic Games, Valve and Ikea).
XR refers to extended reality --- encompassing all of Milgram's reality--virtuality continuum~\cite{milgramAugmentedRealityClass1994}.
From Unity, with an OpenXR target, the project is built into a Visual Studio project.


Visual Studio is Microsoft's IDE for Microsoft software platforms like Win32 API, .NET and UWP.
To build the project in Visual Studio, the target architecture is ARM64, as the HoloLens contains an ARM Qualcomm Snapdragon 850~\cite{HoloLens}.
Once the target architecture is set, and the startup project is configured, the game can be built and run on the HoloLens.
Visual Studio supports this over the network or through a USB-C cable.
Figure~\ref{fig-vs-dev} shows the configuration of the IP address for the remote execution of a project for the HoloLens;
figure~\ref{fig-hololens-mr-dev} shows an example of developing in mixed reality using a USB-C cable.
We sideloaded the game onto the headset through this process.
Once installed the program can be played offline.
Note that this is `compiled' and not `interpreted'; it is not possible to develop live on the headset currently.

In summary, the development cycle is a 2 step process, consisting of building a Unity project to a Visual Studio project, and building from the Visual Studio project to the HoloLens.
While the development process is primarily done through GUIs and is supported platforms are constrained to Microsoft's ecosystem, we can say that {\itshape `We have the hardware\dots to support low latency augmented reality interactions'}.

\subsection{Mixed Reality Control}
\label{sec-mixed-reality-control}

\begin{figure}[h!]
\begin{framedlisting}[left skip=15pt,right skip=15pt]{language=c, numbers=left, escapechar=\&}{C\hash}
float targetX = camera.localPosition.x * headsetMovementMultiplication&\\& + targetXOffset;
float xDiff = targetX - transform.position.x;
// exponential velocity curve
float xVelocity = (float) (1 - Math.Pow(xDiff / maxXDiff + 1, -2)) * maxXVelocity;
Vector3 velocity = body.velocity;
velocity.x = xVelocity + horizontal * maxXVelocityController;
body.velocity = velocity;

// If controller other than headset being used add offset to compensate
if (horizontal != 0) {
	targetXOffset = transform.position.x;
}

Vector3 cameraPosition = transform.position + cameraOffset;
cameraPosition.x = targetX - camera.localPosition.x;
camera.parent.position = cameraPosition;
\end{framedlisting}
\caption{Movement code for the HoloLens Cubes game using the MRTK APIs.}
\label{fig-hololens-movement-code}
\end{figure}

There are a number of interaction models supported by MRTK, including using hands and motions, voice, or `gazing' (interacting with the user's eyes).
While all of these will be extremely useful in providing an intuitive interface for the AR interface, they are unnecessary for our game.
Previously, the game was controlled through a keyboard;
the player used the arrow keys to move the blue player cube.
Touch controls were also included for mobile devices.
However, neither of these control mechanisms is suited to AR.
Instead, a natural control method for a user with a headset is the headset's location.
The MRTK API exposes a camera object representing the headset, which is updated in the game engine as the user moves.
This can be accessed by a script in order to move the player object accordingly.
Figure~\ref{fig-hololens-movement-code} shows MR-specific movement code for the player object.

Line 1 sets the target X position of the player to the headset position is obtained from \texttt{camera.localPosition} with \texttt{camera} being set to the appropriate component from the Unity editor.
The movement of the headset was scaled down by a factor of 20 with \texttt{headsetMovementMultiplication} as the game world operates in meters squared cubes, which turned out to be too far for to player to move to react in time to oncoming cubes.
With the game's platform, the white base visible in figure~\ref{fig-unity-dev}, having a width of 15 meters, this means the user moves within a 75cm area.

The \texttt{targetXOffset} is used to support control mechanisms other than the location of the headset, for debugging.
This could be a keyboard, or gamepad, connected to the headset via Bluetooth.
We need to apply an offset based on this input or else the \texttt{targetX} will always be entirely be determined by the headset's location.
We set the value of \texttt{targetX} in lines 9 -- 12 based on horizontal input obtained from:
\begin{framedlisting}{language=c, tabsize=4}{C\hash}
void Update() {
	horizontal = Input.GetAxis("Horizontal");
}
\end{framedlisting}
This is separated from the previous code, which is in the \texttt{Update} method, as the \texttt{Update} method is run at a fixed rate once per frame, whereas the \texttt{FixedUpdate} is called at fixed intervals independent of the game's framerate.
Conflating these can lead to strange results.

Once we have the \texttt{targetX} position, we don't move the player object there directly, as the location of the camera can be rather jerky, it felt unnatural to have the cube follow the player directly.
Instead, we calculate the distance between where the player is (\texttt{transform.position.x}) and where they are going \texttt{targetX} as \texttt{xDiff}.
On line 4 we then calculate an exponential velocity curve to move the player towards that location:
$$\texttt{xVelocity}=\left (1-\left (\frac{\texttt{xDiff}}{\texttt{maxXDiff}}+ 1\right )^2 \right )* \texttt{maxXVelocityController}$$
We apply this velocity along with the input from another controller, \texttt{horizontal * maxXVelocityController}, on lines 5--6.

Lines 14--16 update the camera position to support scaling the player movement down with \texttt{headsetMovementMultiplication} by applying an offset to the \texttt{camera} object's parent.

\clearpage
\subsection{Holographic Terminal}
\label{sec-holographic-terminal}

To explore lower level interaction with the HoloLens, like using mDNS resolution, we attempted to get a command line interface on the device.
Combined with a Bluetooth keyboard this could be used to interact with ubicomp devices instead of using polished gesture controls, of potentially great interest to the technically competent and for operations that fall outside standard use cases.
The Windows Terminal is a modern terminal emulator that supports the Windows command prompt, powershell and Windows Subsystem for Linux~\cite{WelcomeWindowsTerminal2022}.
The de-factor standard terminal emulator for Windows was a natural choice to use for the HoloLense 2.
And the original HoloLens supported the Windows Terminal~\cite{WindowsTerminalArchitecture}.

However, one cannot install the application from the Windows store, as the application's target platform is \texttt{Windows.Desktop}.
After trying and failing to get the Windows Terminal built locally~\cite{ErrorWhenBuilding}, we used the MSIX Packaging Tool~\cite{GetMSIXPackaging} to change the target platform to \texttt{Windows.Desktop}.
However, after sideloading the \texttt{MSIX} file, launching it on the HoloLens failed.
It turns as the HoloLens uses an ARM architecture the Win32 API isn't supported.
The Windows terminal uses XAML islands to support UWP, but this is really just for graphical components to ensure uniformity on the new Windows versions and it still uses Win32 under the hood~\cite{HololensBuildIssue}.
The original HoloLens supported the Win32 API and by extension the Windows Terminal as it contained an x86 Intel Cherry Trail SoC.

\begin{figure}[b!]
	\centering
	\includegraphics[width=1\textwidth]{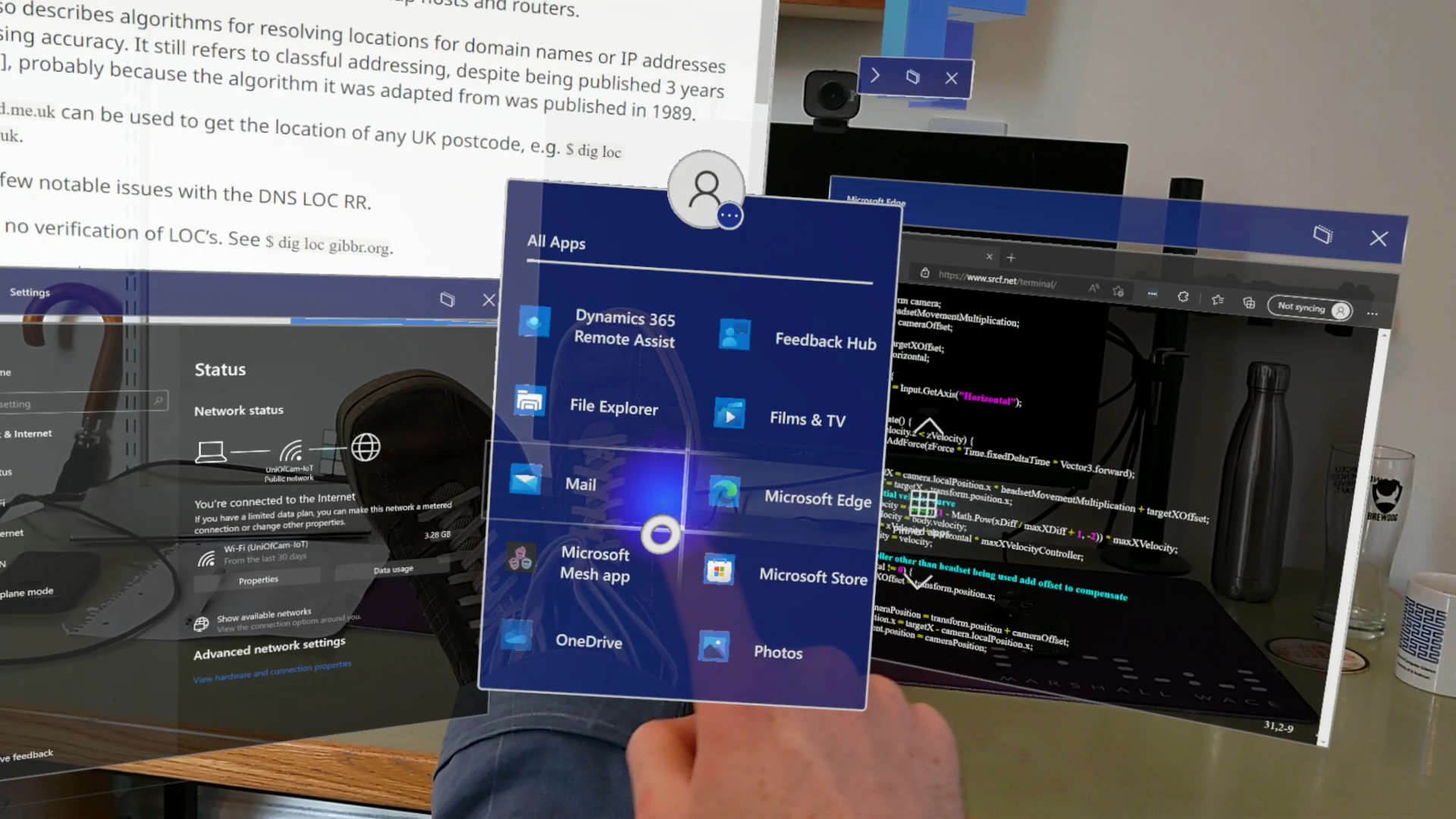}
	\caption[]{A view of overlaid virtual windows from the HoloLens.}
	\label{fig-hololens-windows-view}
\end{figure}

One can use a remote shell through the included Edge browser, like the Cambridge Student Run Computing Facility's (SRCF) hosted Shell In A Box.
A Bluetooth keyboard is useful to avoid having to type on a holographic keyboard --- which is very futuristic, but also very slow.
Figure~\ref{fig-hololens-windows-view} shows an example of window management on the HoloLens with this shell. 
The window management system is very impressive, with gesture control being used to move and pin windows, despite being significantly less efficient and error-prone than traditional peripherals or touchscreens.
There is an anchor system that provides an OS abstraction that persists windows and other objects in a room that the HoloLens recognises.
The device portal API also proved very useful in watching a real-time feed of the user's POV and providing help~\cite{hamalawiDevicePortalAPI}.
\todo{this}

\FloatBarrier

\section{Interactivity Evaluation}
\label{sec-interactivity-evaluation}

We have shown that {\itshape `We have the hardware and software to support low latency augmented reality interaction'} in the form of a cutting edge AR headset, a development environment for it, and tookits such as spatial mapping, hand tracking and interaction APIs.
However, we referred to the control mechanism we created as mixed reality control~(\S\ref{sec-mixed-reality-control}).

\begin{figure}[b!]
	\centering
	\includegraphics[width=0.5\textwidth]{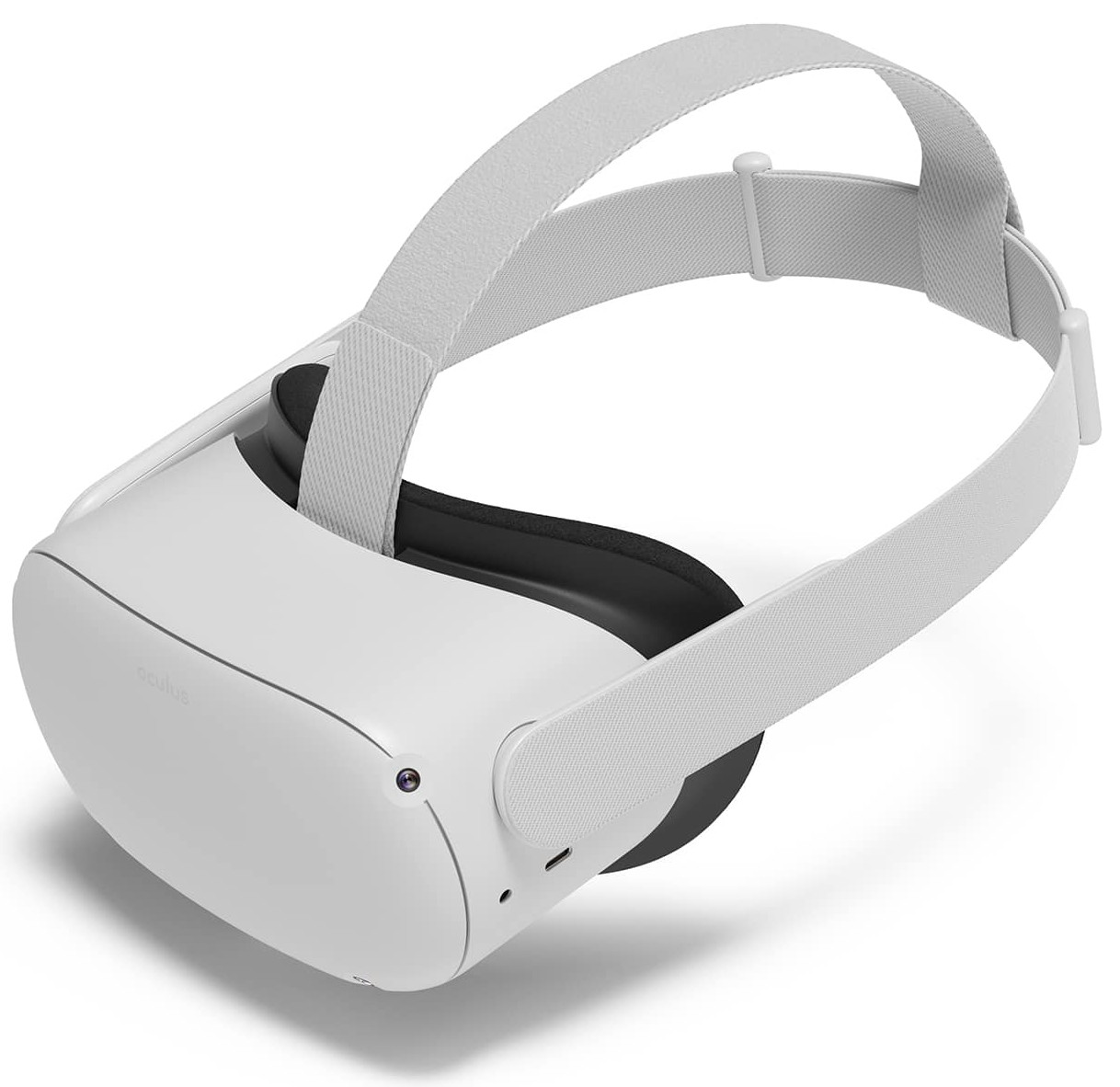}
	\caption[]{Meta's virtual reality headset Quest 2~\cite{QuestOurMost}}
	\label{fig-quest-2}
\end{figure}

This is because the mechanism does not fundamentally change between AR and VR, it just uses different APIs.
We discovered this from porting the game to the Meta Quest 2 VR headset, shown in figure~\ref{fig-quest-2}.
The only difference between our game in VR and AR is that in the latter one can see the real world in the background.
This allows one to, say, drink a mug of coffee more easily while wearing it but their interaction models are fundamentally the same.

However, we said that we want to use this headset for AR rather than MR~(\S\ref{sec-hw-and-sw}).
This brings us to the primary limitation in the interactivity of our game - it has not met our goal of interacting with the real physical world.

There are spatial mapping APIs available for the HoloLens that allow us to overlay virtual objects with knowledge of a room's physicality topology.
For example, there is an API to find good places to render virtual objects.
But still, this is constrained to an overlay of virtual objects that aren't connected to the real world and is not fulfilling the full potential of AR.
Looking around the Microsoft store for applications for the HoloLens, they are generally constrained to VR games overlaying the real world and 3D rendering viewing of models, but nothing you can't do better on a VR headset.
To put this in the context of Milgram's reality--virtuality continuum, these applications are low on the Extent of World Knowledge (EWK) spectrum~\cite{milgramAugmentedRealityClass1994}.
Weiser referred to ubicomp as embodied virtuality, to contrast it with virtual reality~~(\S\ref{sec-ubicomp}).
However, the MX of the HoloLens is more similar to VR than this vision.
There are interesting examples of applications that go beyond this, like remote maintenance assistance, and there are many visions of tying this into the real world, such as for monitoring industrial equipment.
But there is no killer app for AR yet.


Interaction with computation devices in the real world could be that killer app.
While there are numerous spatial APIs available with the HoloLens's advanced array of sensors, there is little such support for networking.
Unity has support for networking games for multiplayer.
This includes a high level multiplayer API, but also a `low level' transport layer API that is implemented as a shim about OS sockets~\cite{TransportsUnityMultiplayer}.
But there is limitted support for local networking.
It is possible to manually use UDP multicast for discovery, designed for connecting games on a local network~\cite{MigratingUNetNetcode}.
But it is obviously undesirable to run Unity on resource constrained ubicomp devices, and this does not provide a naming mechanism for physical devices.
If we wanted to use mDNS, despite its issues~~(\S\ref{sec-internet-arch}), we would have to re-implement it in Unity.
We might think mDNS support is something that could be added to the OS through a userland package or kernel module, and windows do indeed provide a DS-DNS (Discovery Service DNS - implemented on top of mDNS) API for the Win32 API;
but as we have discovered trying to build the Windows Terminal for the HoloLens it does not support this API due to its ARM architecture~\cite{stevewhimsWindowsNetworkingServiceDiscovery}.
We could use the existing DNS infrastructure trivially with the Unity transport layer API, but it is inadequate for our low latency, reliable, secure and private ubicomp use case~~(\S\ref{sec-internet-arch}).


In summary, what we want to do is build a local AR application that can interface with devices in the local environment.
But the systems support required for this is just not there.
The Microsoft and game development ecosystem is not easy to do low level systems work in.
If we want these devices to succeed for use cases beyond those already existing we need them to be as free and as open as possible, which is especially true as they become part of our physical infrastructure and environment.
However, the more pressing issue is that the network support is inadequate.
We have shown that we can build low latency interaction applications locally on the HoloLens, but no appropriate naming mechanism for the physical world.
Therefore, {\itshape `We need a Spatial Name System that can map physical devices to network addresses to overcome this limitation and unlock the potential of augmented reality.'}
We will explore this system in chapter~\ref{ch-spatial-networking}.

\section{Summary}
\label{sec-augmented-reality-interface-summary}

In this chapter, we started by selected cutting edge augmented reality hardware and software~(\S\ref{sec-hw-and-sw}).
We then explored the development environment for augmented reality~(\S\ref{sec-ar-development}) using a Unity development environment to create a low latency game~(\S\ref{sec-development-process}) with mixed reality controls~(\S\ref{sec-mixed-reality-control}), and recounted our experience trying to hack a holographic terminal onto the HoloLens~(\S\ref{sec-holographic-terminal}).
We conclude with an evaluation of the limitations we encountered trying to build an interactive interface with networking on HoloLens that prevent us from realising our vision of an Augmented Reality Interface and justify our Spatial Name System~(\S\ref{sec-interactivity-evaluation}).

\chapter{Spatial Networking}
\label{ch-spatial-networking}


In this chapter, we will test the thesis {\itshape `We need a Spatial Name System that can map physical device locations to network addresses to overcome this limitation and unlock the potential of augmented reality'} by proposing a network architecture to map physical device locations to network addresses.
The `Spatial Name System' (SNS) will overcome the limitations of networking AR with the physical world and unlock the potential of augmented reality~(\S\ref{sec-interactivity-evaluation}).
In order for this system to work, we need to know the know local positions of ubicomp devices.
These locations can be obtained by using a beacon with a local positioning system~(\S\ref{sec-local-positioning-systems}) or for non-mobile devices by statically surveying them~(\S\ref{sec-mobility}).
Once we know where devices are locationed, we need a system to map them to network addresses.
We begin by describing the technologies used towards achieving this end~(\S\ref{sec-technologies}), then explore a DNS resource record used to encode location~(\S\ref{sec-loc-rr}), we will consider a spatial mapping system to resolve physical locations to network addresses~(\S\ref{sec-spatial-resolution}), and propose an architecture for creating this system.

\section{Technologies}
\label{sec-technologies}

Traditionally C has dominated systems programming.
In the last 2 decades, we have seen an increase in higher-level languages being used for lower-level programming, such as Rust.
As our SNS will be a critical part of the infrastructure in a context where security and privacy are of the utmost importance, we care a lot about correctness and safeness.
For this reason, we will use the functional programming language OCaml, which comes with a number of benefits~\cite{madhavapeddyMelangeCreatingFunctional2007,madhavapeddyUnikernelsLibraryOperating2013}.
\begin{enumerate*}[(1)]
	\item Static type checking classifies functions and variables into types and catches errors in the operations on these types at compile time instead of run time.
	A language like C does not have as rich a type system, relying on primitive types, structs and unions.
	Even arrays are simply pointers into memory;
	the programmer has to make sure the data is being interpreted in the correct way.
	\item Garbage collection automatically manages memory preventing memory leaks.
	Techniques like incremental and generational collection minimise the performance impact of interruptions to the application.
	Manual memory management can still be done when required.
	\item Modules and object-orientated features allow composable components with well-defined interfaces.
\end{enumerate*}


We start our work towards the Spatial Name System by exploring the existing support for location-based DNS in section~\ref{sec-loc-rr}.
This was done in the OCaml DNS library, created as part of the MirageOS project~\cite{OpamDns, madhavapeddyUnikernelsLibraryOperating2013}.
In addition to the benefits gained from using OCaml, this DNS stack is a lot more malleable than the alternatives.
The most common known DNS server on the Internet by far is Bind9~\cite{sissonDnsSurvey20102010, consortiumBIND}.
Bind is written in C and dates from the early 1980s.
While lines-of-code (LOC) is not a perfect metric, and functional code tends to be very expressive, the OCaml DNS implementation is \textit{significantly} smaller than the Bind9 server.
The OCaml DNS server is 76,735 LOC compared to Bind9's 373,578 LOC.
This is 20\% the LOC of Bind9.
Note the LOC count included header and interface files, and includes all libraries installed by the OCaml package manager \texttt{opam}.
Excluding these libraries, the OCaml DNS implementation is only 2,471 LOC.
Additionally, the OCaml modules system with well-defined interfaces allows existing code bases to be modified much more conveniently than C.

\begin{framedlisting}{upquote=true}{Bash}

$ find Bind9 -type f \( -name '*.c' -o -name '*.h' \) -exec cat {} + | wc -l
373578

$ find ocaml-dns -type f \( -name '*.ml' -o -name '*.mli' \) -exec cat {} + | wc -l
76735
\end{framedlisting}

We mentioned that OCaml DNS was created as part of the MirageOS project~\cite{madhavapeddyUnikernelsLibraryOperating2013}.
Mirage is a library operating system that creates a small single address space operating system targetting a specific architecture.
This has benefits for safety and security, with the whole kernel being type-checked.
Dead code elimination means that built unikernels are much smaller than an application dragging along an entire operating system with it.
Unikernels are compiled alongside their configuration, which means with metaprogramming the compiler can optimize the application significantly.
While mirage was originally used to target virtualization environments like the Xen hypervisor~\cite{barhamXenArtVirtualization2003}, it can also target bare-metal embedded systems including the Broadcom BCM2837 and ESP32 Chip.
Due to all these benefits, Mirage can benefit programming for ubicomp and embedded devices immensely.

\section{LOC Resource Record}
\label{sec-loc-rr}

Similar to how ARP resolves network addresses to MAC addresses, and DNS resolves domain names to network addresses, SNS resolves locations to network addresses.
While DNS was original simply for resolving domain names to network addresses, its purpose has expanded over the years with various resource records, such as MX records for mailservers, SRV for generic service discovery, and PTR records for resolving IP addresses to domain names.
DNS contains a mechanism for associating a domain name with a physical location, the `LOC' resource record (RR).
Note this is LOC as in location, not lines-of-code.

The LOC record was proposed in RFC 1876 `A Means for Expressing Location Information in the Domain Name System'~\cite{dickinsonMeansExpressingLocation1996}, as a refinement of the earlier GPOS record from RFC 1712 `DNS Encoding of Geographical Location'~\cite{schulzeDNSEncodingGeographical1994}.
It encodes latitude, longitude, altitude, size of a referenced sphere, horizontal precision and vertical precision.


There are a number of suggested uses proposed for the LOC record that shows its RFC's age.
They include USENET geographic flow maps; a 'visual traceroute' application showing the geographical flow of IP packets, relying on routers responding to both IP TTL timeouts and DNS LOC requests; and network management based using LOC RRs to map hosts and routers.
These visions have not been realised in practice, for security and privacy reasons, as well as a lack of a real use case.
They have limited real-world usage and no practical use that we are aware of, generally relegated to easter eggs~\cite{WeirdWonderfulWorld}.

However, our Spatial Name System is a practical case for using location information in a naming system.
While the LOC record is a mechanism for global positioning rather than local positioning, using this existing record could save adding an additional location encoding resource record.
For these reasons, we implemented this record in the OCaml DNS library.

First, we added a new module \texttt{LOC} to \verb|src/dns.mli| (figure~\ref{fig-loc-module}).
\begin{figure}[h]
\begin{framedlisting}[left skip=15pt,right skip=15pt]{language=ml,escapechar=\&}{OCaml}
module Loc : sig
	type t = {
		latitude : int32;
		longitude : int32;
		altitude : int32;
		size : int;
		horiz_pre : int;
		vert_pre : int;
	}
	(** The type of a Loc record. *)
	&\ldots&
end
\end{framedlisting}
\caption{\texttt{Loc} module type for LOC resource record.}
\label{fig-loc-module}
\end{figure}

With OCaml's module system and non-exhaustive pattern matching check, we could then find all the cases where we needed to modify the code base.
Most of these cases were trivial but some required special treatment, such as printing the record:
\begin{framedlisting}{language=ml}{OCaml}
| Loc, (ttl, locs) ->
	Loc_set.fold (fun loc acc ->
	Fmt.str "%s\t%aLOC\t%s" str_name ttl_fmt (ttl_opt ttl) (Loc.to_string loc) :: acc)
	`locs []
\end{framedlisting}

\clearpage
\subsection{Parsing}
\label{sec-parsing}

DNS records are generally created by listing them in a zone file associated with a zone.
These records can be for the zone origin, or any subdomain in the zone.
The LOC resource record has a relatively complicated zone file format:
\begin{center}
\begin{tabular}{c}
\begin{lstlisting}
The LOC record is expressed in a master file in the following format:

<owner> <TTL> <class> LOC ( d1 [m1 [s1]] {"N"|"S"} d2 [m2 [s2]]
                            {"E"|"W"} alt["m"] [siz["m"] [hp["m"]
                            [vp["m"]]]] )

(The parentheses are used for multi-line data as specified in [RFC
1035] section 5.1.)

where:

    d1:     [0 .. 90]            (degrees latitude)
    d2:     [0 .. 180]           (degrees longitude)
    m1, m2: [0 .. 59]            (minutes latitude/longitude)
    s1, s2: [0 .. 59.999]        (seconds latitude/longitude)
    alt:    [-100000.00 .. 42849672.95] BY .01 (altitude in meters)
    sizz, hp, vp: [0 .. 90000000.00] (size/precision in meters)
\end{lstlisting}
\end{tabular}
\end{center}

\noindent
An example LOC record value is:
\begin{center}
	\begin{tabular}{c}
		\begin{lstlisting}
52 12 40.4 N 0 5 31.9 E 22m 10m 10m 10m
\end{lstlisting}
	\end{tabular}
\end{center}

The C code that the RFC helpfully provides lexes, parses and encodes this in one step.
The lexing consists of splitting the input into tokens; the parsing extracts values from these tokens according to their syntax; and the encoding consists of converting the human-readable representation into a format for transmission on the wire.
While this is efficient, it is also less composable and more error-prone than separating these steps.
We will use the libraries \texttt{ocamllex} and \texttt{ocamlyacc}, which replace the venerable \texttt{lex} and \texttt{yacc} respectively, to implement this parsing in OCaml.
Functional languages are well suited to the task of parsing due to features like pattern-matching and good support for recursion.

Parsing can be described as analysing a string of tokens according to the rules of a formal grammar.
We can describe the LOC record format as a context-free grammar using an extended BNF notation~\cite{backusRevisedReportAlgorithm1963}.
In the extended syntax, we represent terminals as \textsl{term}, literals as {\ttfamily\bfseries literal}, alternation with \{\textsl{term1} $|$ \textsl{term2}\}, optional elements as [\textsl{optional}], elements which must occur one or more times as (\textsl{term})\verb|+|, and whitespace consisting of spaces and tabs with $<${\ttfamily\bfseries space}$>$.

\begin{center}
	\begin{tabular}{rl}
		\slshape type\_loc \verb|::=|
		& $<${\ttfamily\bfseries space}$>$ \textsl{deg\_min\_sec} \{{\ttfamily\bfseries N} $|$ {\ttfamily\bfseries S}\} \\
		& $<${\ttfamily\bfseries space}$>$ \textsl{deg\_min\_sec} \{{\ttfamily\bfseries E} $|$ {\ttfamily\bfseries W}\} \\
		& $<${\ttfamily\bfseries space}$>$ \textsl{meters} \\
		& $<${\ttfamily\bfseries space}$>$ \textsl{meters} \\
		& [$<${\ttfamily\bfseries space}$>$ \textsl{meters} [$<${\ttfamily\bfseries space}$>$ \textsl{meters}]] \texttt{;} \\

		\slshape deg\_min\_sec \verb|::=|
		& \textsl{integer} [$<${\ttfamily\bfseries space}$>$ \textsl{integer} [$<${\ttfamily\bfseries space}$>$ \textsl{integer} [{\ttfamily\bfseries .} [\textsl{integer}]]]] \texttt{;} \\

		\slshape precision \verb|::=|
		& \textsl{meters} \\

		\slshape meters \verb|::=|
		& [{\ttfamily\bfseries -}] \textsl{integer} [{\ttfamily\bfseries .} [\textsl{integer}]] [{\ttfamily\bfseries m}] \texttt{;} \\

		\slshape integer \verb|::=| & ({\ttfamily\bfseries 0} $|$ {\ttfamily\bfseries 1} $|$ {\ttfamily\bfseries 2} $|$ {\ttfamily\bfseries 3} $|$ {\ttfamily\bfseries 4} $|$ {\ttfamily\bfseries 5} $|$ {\ttfamily\bfseries 6} $|$ {\ttfamily\bfseries 7} $|$ {\ttfamily\bfseries 8} $|$ {\ttfamily\bfseries 9})\texttt{+} \texttt{;} \\
	\end{tabular}
\end{center}

In practice, it was more complicated to implement than this grammar would suggest.
One issue was having to interact with the existing parsing for the rest of the records.
Using {\ttfamily\bfseries m} as a token proved problematic.
Take an example instance of the \textsl{meters} term as \verb|0m|.
This can be lexed as an \textsl{integer} and {\ttfamily\bfseries m}, but it can also be lexed as a generic \textsl{charstring} {\ttfamily\bfseries 0m}.
The way \texttt{ocamllex} deals with ambiguity is by taking the longest match as a winner.
In this case, that would be the \textsl{charstring}.
We could modify the \textsl{charstring} to ignore strings beginning with a digit, but then we couldn't tokenise a generic string starting with an integer.
This would break any domains starting with a number~\cite{bradenRequirementsInternetHosts1989}.

Instead, we used a regex to tokenise the whole \textsl{meters} term at once.
However, the \texttt{ocamllex}'s ambiguity resolution foiled us again.
IPv4 addresses are tokenized as a series of integers and dots ({\ttfamily\bfseries .});
for example, \verb|192.0.2.0|.
As the longest match wins, the \textsl{meters} term with an optional {\ttfamily\bfseries m} will match \verb|192.0| and break the lexing for the IP address.
To resolve this we took a hybrid approach between lexing and parsing.
We tokenized:
\vspace{-1em}
\begin{center}
	\begin{tabular}{rl}
		\textsl{meters} \verb|::=| [{\ttfamily\bfseries -}] \textsl{integer} [{\ttfamily\bfseries .} [\textsl{integer}]] {\ttfamily\bfseries m} \texttt{;}
	\end{tabular}
\end{center}
\vspace{-1em}
Note the mandatory {\ttfamily\bfseries m}).
The other cases, where {\ttfamily\bfseries m} doesn't occur, were matched in the parser by using \textsl{integer}, {\ttfamily\bfseries .}, and {\ttfamily\bfseries -} tokens.

Once we had these tokens, we had to evaluate them as numeric datatypes.
We wanted to avoid using any floating point numbers as they can be imprecise, require converting to another representation for packing them on the wire, and there are issues using them in-kernel as MirageOS unikernels do: how they are handled varies by architecture and it requires enabling the floating-point unit.
Instead, we parsed the float to the number of decimal places specified in the RFC: 3 decimal places for seconds latitude/longitude, and centimetre accuracy (2 decimal places) for meter measurements.

\clearpage
\subsection{Encoding}
\label{sec-encoding}

Once we lexed and parsed the input, we then needed to translate it to the on the wire format.
The format of the record is:
\begin{center}
\begin{tabular}{c}
\begin{lstlisting}
     MSB   LSB
    +--+--+--+--+--+--+--+--+--+--+--+--+--+--+--+--+
   0|        VERSION        |         SIZE          |
    +--+--+--+--+--+--+--+--+--+--+--+--+--+--+--+--+
   2|       HORIZ PRE       |       VERT PRE        |
    +--+--+--+--+--+--+--+--+--+--+--+--+--+--+--+--+
   4|                   LATITUDE                    |
    +--+--+--+--+--+--+--+--+--+--+--+--+--+--+--+--+
   6|                   LATITUDE                    |
    +--+--+--+--+--+--+--+--+--+--+--+--+--+--+--+--+
   8|                   LONGITUDE                   |
    +--+--+--+--+--+--+--+--+--+--+--+--+--+--+--+--+
  10|                   LONGITUDE                   |
    +--+--+--+--+--+--+--+--+--+--+--+--+--+--+--+--+
  12|                   ALTITUDE                    |
    +--+--+--+--+--+--+--+--+--+--+--+--+--+--+--+--+
  14|                   ALTITUDE                    |
    +--+--+--+--+--+--+--+--+--+--+--+--+--+--+--+--+
(octet)
\end{lstlisting}
	\end{tabular}
\end{center}
Where one line is 2 octets, where an octet is 8 bits, and the overall size is 16 bytes if we consider a byte to also be 8 bits.
The \verb|SIZE|, \verb|HORIZ PRE|, and \verb|VERT PRE| fields are only 8 bits each but represent values up to $9*10^9$cm.
This is done by using the first 4 bits as a mantissa (the base), and the next four bits as the exponent (the power of 10 by which to multiply the base).
OCaml's strong tying and boxed types proved invaluable for this encoding, but this did required implementing integer exponentiation recursively:
\begin{framedlisting}{language=ml}{OCaml}
let exponent =
	let rec r = fun p e ->
		if e >= 9 then 9 else
		if p < (pow10 (e + 1)) then e else
		r p (e + 1)
	in
	r p 0
\end{framedlisting}

To encode the latitude and longitude, the sexagesimal (base 60) degrees - degrees, minutes and seconds - had to be converted to thousandths of a second of arc; that is, milliarcseconds~(mas).
The on the wire format was then simply the 32-bit representation of this integer:
\begin{framedlisting}{language=ml}{OCaml}
let lat_long_parse ((deg, min, sec), dir) =
	let ( * ), (+) = Int32.mul, Int32.add in
	(Int32.shift_left 1l 31) + (
		(((deg * 60l) + min) * 60l) * 1000l + sec
	) * if dir then 1l else -1l
\end{framedlisting}
Conversion to decimal degrees can be done by multiplying this milliarcsecond value by $60*60*1000$.



Finally, to encode the altitude, we simply subtracted 100000m from it, in order to represent the range of values -100000.00 -- 42849672.95m.
Note the maximum value represented by an unsigned 32-bit integer is
$$4284967295+10000000=4294967295=2^{32}-1$$
This means we can represent objects at an altitude of 40,000km.
For reference, the International Space Station orbits at 400km.

While encoding the altitude was the simplest of all the fields, decoding it was the most troublesome.
OCaml does not have inbuilt support for unsigned types.
There is an \texttt{ocaml-integers} library providing this functionality, but it requires adding another dependency and pulls a fair amount of C along with it.
Instead, we manually converted it to an \texttt{int64} by compensating for the subtraction of $2^{32}$ if the 32nd bit is 1 when interpreting an unsigned 32-bit integer as a two's compliment number.
\begin{framedlisting}{language=ml}{OCaml}
(* convert a uint32 alt to an int64 *)
let alt = if alt < 0l then
		Int64.of_int32 alt + Int64.shift_left 1L 32
	else Int64.of_int32 alt
\end{framedlisting}

\subsection{Testing}
\label{sec-testing}

We performed extensive testing on the LOC record's parsing, printing, encoding and decoding.
To test parsing and printing, test records were printed and compared to what was expected.
The expected printing wasn't always what was parsed due to optional terms or if redundant precision was given, like \verb|0.00| altitude being printed as \verb|0m|.
Notice from the internal representation of the \texttt{Loc} module (figure~\ref{fig-loc-module}) that all the values are stored in their on the wire format.
We did this to ensure what was printed from a record was the same as what will be transmitted on the wire.

To test the packet decoding, the binary format of a LOC record query was dissected and annotated (figure~\ref{fig-dns-loc-query-answer}).
We checked compared against reference binary formats obtained from \texttt{\$ tcpdump -X -n port 53} and using \texttt{dig loc <existing domain>} where \texttt{<existing domain>} is a domain controlled by the author running \texttt{Bind9}.
In an indirect way, this is essentially reverse engineering \texttt{Bind9}.

While the RFC's textual description was helpful, it provided invaluable to read the C code to get the exact details of how to encode fields.
Even still, mistakes were made that were only caught by extensive testing.
There is work towards formalising RFCs in the proof assistant Coq, which could have proved very useful in this case~\cite{affeldtFormalNetworkPacket2012}.

\begin{figure}[h]
\begin{framedlisting}{language=ml}{OCaml}
(* RFC1035 section 4.1 *)
(* header *)
"11 11 81 80 00 01 00 01 00 00 00 00" ^
(* question *)
	(* example.com *)
	"07 65 78 61 6d 70 6c 65" ^ (* example *)
	"03" ^ (* dot *)
	"63 6f 6d" ^ (* com *)
	"00" ^ (* \0 - null terminated *)
	(* QTYPE = 29 = LOC *)
	"00 1d" ^
	(* QCLASS = IN `*)
	"00 01" ^
(* answer *)
	"c0 0c" ^
	(* binary: 11000000 00001100
	first 2 11's indicate a pointer
	000000 00001100 = 12
	this gives the size of the header as an offer, pointing to the domain in the query
	*)
	"00 1d" ^ (* TYPE = LOC *)
	"00 01" ^ (* CLASS = IN *)
	"00 00 0e 10" ^ (* TTL = 3600s *)
	"00 10" (* RDLENGTH = 16 *)
	\end{framedlisting}
	\caption{Anatomy of a DNS LOC query answer.}
	\label{fig-dns-loc-query-answer}
\end{figure}

\FloatBarrier
\section{Spatial Resolution}
\label{sec-spatial-resolution}

With our newly implemented DNS LOC record, we can resolve domain names to physical locations.
But there are two problems with this record for our use case:
\begin{enumerate*}[(1)]
	\item it gives a global position, not a local position;
	\item it resolves domain names to physical locations, but we want to resolve physical locations to network addresses.
\end{enumerate*}

Regarding point (1), our highest precision is 0.001 arcseconds, or 1 milliarcsecond, which corresponds to approximately $2.7*10^{-7}$ decimal degrees.
This is precise to approximately 30 centimetres at the equator and about 10 centimetres at 67\textdegree N.
This is just outside the centimetre precision we would want for small ubicomp devices.
Additionally, with ubicomp devices, we may not know the global position of devices but just their local relative positions.
We could `hack' the record to sacrifice global positioning for higher precision local positioning, reinterpreting latitude and longitude as x and y coordinates;
as at the local scale, the curvature of the Earth won't affect our resolution.
But then we would then be constrained to using a record ill-suited to our needs and would break compatibility with other systems using the LOC record anyway.
\clearpage
To address point (2), the PTR record can be used to perform a reverse DNS lookup, resolving an IP address to a domain name~\cite{DomainNamesImplementation1987, lawrenceObsoletingIQUERY2002}.
This is done with the \verb|IN-ADDR.ARPA| domain.
For example, \verb|0.2.0.192.IN-ADDR.ARPA| should resolve to the domain(s) that point to \verb|192.0.2.0|.
Conceivably, we could do a similar reverse lookup to associate a location with a domain name, which could then be resolved to a network address as normal.
However, using DNS would have implications for latency, reliability, security and privacy~(\S\ref{sec-dns}).
Instead, we could try and modify mDNS to support this lookup.
This would have the advantage that we wouldn't need to worry about modifying remote server's DNS stacks, we could just update end-host ubicomp devices.
And it would address our issue with mDNS, {\itshape `names are descriptive strings, and name conflict resolution relies on programmatic updates, which doesn't provide a reliable or intuitive interface for ubicomp devices'}~(\S\ref{sec-dns}).
However, we shall consider this more in section~\ref{sec-spatial-arch}.


Instead of using the DNS LOC record, we propose an alternative spatial resolution mechanism to resolve a physical location to a network address.
Recall that this system is being designed as a mechanism for an AR headset~(\S\ref{sec-augmented-reality}) to address the world of ubicomp devices.
The headset will know its location through spatial tracking and will know where the user is interacting through gaze, gestures or voice.
This may come with some degree of precision depending on the APIs.
Recall that Weiser identified location and scale as having crucial importance for ubicomp devices~(\S\ref{sec-ubicomp}).
We will conflate scale and precision into a search area for the sake of simplicity, but it could be possible to have richer APIs to distinguish these.

\subsection{Spatial Mapping}
\label{sec-spatial-mapping}

\begin{figure}[t!]
	\centering
	\includegraphics[width=0.65\textwidth]{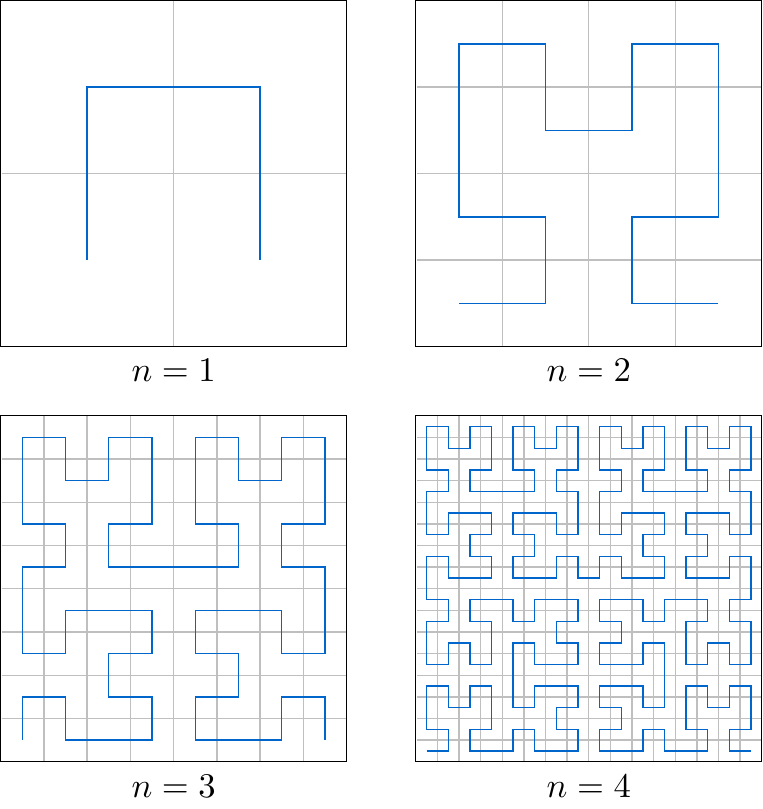}
	\caption{Hilbert Curves with order $n$~\cite{hilbertUeberStetigeAbbildung1891}}
	\label{fig-hilbert-curves}
\end{figure}

The first problem as part of this spatial resolution mechanism is mapping a higher dimensional space into a lower-dimensional space.
Note that in this context mapping refers to a prescribed way of assigning each point in our higher dimensional space to a point in our lower-dimensional space, not cartography.
To achieve this we can use Hilbert curves: space-filling curves that fill a 2D square with a 1D line~\cite{hilbertUeberStetigeAbbildung1891}.
They are fractal curves, with arbitrary precision based on their order, which represents how many iterations to perform.
It's possible to use other fractal space filling curves, such as Peano curves~\cite{peanoCourbeQuiRemplit1890} or Murray Polygons~\cite{coleHalftoningDitherEdge2005}, but these underlying principles stay the same.
Figure~\ref{fig-hilbert-curves} diagrams Hilbert curves from order 1 to 4.
We only diagram in 2 dimensions, but Hilbert curves can be extended to 3 dimensions.

\begin{figure}[b!]
	\centering
	\includegraphics[width=0.6\textwidth]{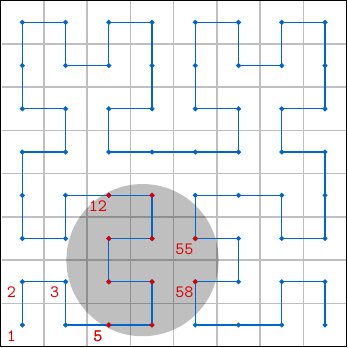}
	\caption{Spatial mapping of the circle in grey to points on a Hilbert curve of order 3. The indices of the Hilbert curve that the circle's area intersect are 5--12, 55 and 58, denoted with red points.}
	\label{fig-spatial-map}
\end{figure}

Hilbert curves preserve locality when mapping a 1D space to a higher dimension.
Two points close to each other in 1D space will be close in the higher dimensional space.
This is useful for our spatial mapping as it means we can represent an area through an interval of the Hilbert curve.
Figure~\ref{fig-spatial-map} shows an example of this, where the shaded circle is the area we want to map, representing an object of that scale, or a point with a certain precision.
The numbers in the grid squares index the Hilbert curve at that point on the curve.

Note that while points in 1D space will appear close to each other in 2D space, the converse is not necessarily true, which is why we can't always represent an area by one continuous interval.
This is demonstrated by the fact that the points we map to on the Hilbert curve in figure~\ref{fig-spatial-map} are 5--12, but also 55 and 58.

The simplest alternative to using a space-filling curve would be a progressive scan across the search area.
That is, converting the points to a 1D value with a formal such as $x+y*width$.
This would only maintain an interval horizontally.
Taking the bottom left of the graph as (0, 0), x as horizontal, and y as vertical, would result in the intervals 11--12, 19--21, 27--29 and 35--36 for figure~\ref{fig-spatial-map}.
Compared to our space-filling curve, this preserves locality much less well, and the additional intervals would have implications for the complexity of our lookup.


\subsection{Resolution Mechanism}
\label{sec-resolution-mechanism}

\begin{figure}[h]
	\centering
	\includegraphics[width=0.6\textwidth]{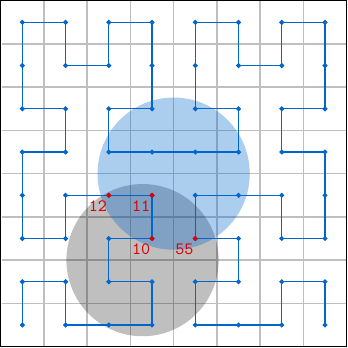}
	\caption{Resolving a spatial location to network address via the overlap of our query area (\textit{grey}) with an address area (\textit{blue}).The indices of the Hilbert curve where the areas overlap are 10--12 and 55.
	}
	\label{fig-spatial-resolution}
\end{figure}

With this mapping from 2 dimensions to 1 dimension, we have the first part of our spatial resolution mechanism.
The second part is how to resolve these intervals to a network address.
We shall describe a \textit{query area} as the intervals associated with an area that a client is querying, and an \textit{address area} as the area associated with a particular network address, also represented by a set of intervals.
We want to find all the address areas that overlap with our query area.
For example, in figure~\ref{fig-spatial-resolution} we can find the overlap of our query area (in grey) with the address area (in blue) as the Hilbert curve indices that are contained in both.

We can use an augmented\footnote{No relation to augmented reality; augmented here refers to augmenting a binary tree.} interval tree to find the overlapping intervals.
An interval tree is a data structure that, given an interval, efficiently finds if it overlaps with any existing intervals.
We use a self-balancing Binary Search Tree (BST) to maintain a set of intervals such that all insertion, deletion, and searching for overlapping intervals can be done in $O(\log n)$ time.
Each node in the tree will contain:
\begin{enumerate}[(1)]
	\item An interval represented by a pair \texttt{[low, high]}.
	\item The maximum value occurring in subtree below the node.
	\item An address which the interval represents the \textit{address area} of.
\end{enumerate}
The low value of a node's interval is used as the BST key.
That is, the invariant that must be maintained is that every node in the left subtree must have its lower interval be less than or equal to the subtree root node, and every node in a right subtree must have its lower interval be greater than the subtree root node.
The self-balancing red-black tree insert and delete operations are excluded for the sake of brevity.

The query logic for this tree is:
\begin{framedlisting}[breakable]{language=c}{}
// check if intervals overlap
overlap(interval1, interval2) {
	return interval1.low <= interval2.high && interval1.high >= interval2.low
}

search(node, interval, addresses) {
	if (node == NULL) {
		return NULL
	}
	if (overlap(node.interval, interval)) {
		addresses.add(node.interval.address)
	}
	if (node.left != NULL && node.left.max >= i.low) {
		search(node.left, interval, addresses)
	}
	else {
		search(node.right, interval, addresses)
	}
}

query(intervals) {
	addresses = set()
	// for every interval in query area
	for interval in intervals {
		search(root, interval, addresses)
	}
	return addresses
}
\end{framedlisting}

\clearpage
This is a modification of a standard augmented interval tree as it returns every overlapping interval, not just the first one found.
This is to deal with overlapping address areas, where we want to return every address found so the client knows about the ambiguity.

\begin{figure}[h]
	\centering
	\includegraphics[width=0.5\textwidth]{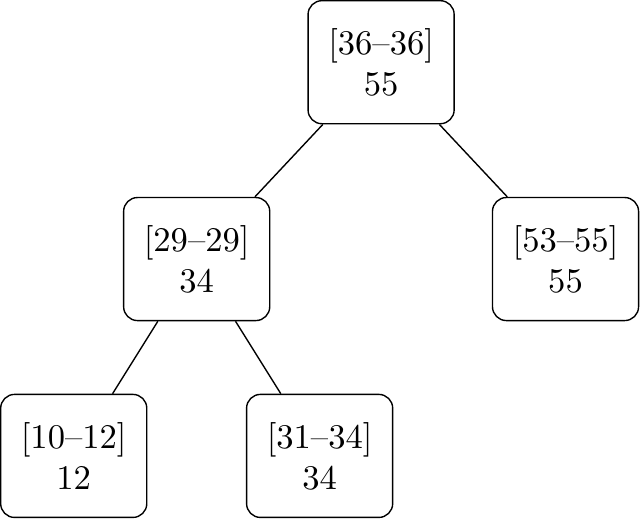}
	\caption{
		The augmented interval tree for the blue address area from figure~\ref{fig-spatial-resolution}.
		The format of the nodes is the interval represented by the pair {[\texttt{low}--\texttt{high}]} with the max value below.
		The address associated with a node is omitted as they all belong to the same address area.
	}
	\label{fig-augmented-tree}
\end{figure}

To illustrate this resolution mechanism, figure~\ref{fig-augmented-tree} illustrates one possible balanced tree associated with the blue address area depicted in figure~\ref{fig-spatial-resolution}.
We shall consider the search on this tree for the first interval from the query area in figure~\ref{fig-spatial-resolution}: [\texttt{5}--\texttt{12}].
When comparing to the root node [36--36] we recurse on the left subtree from the root as $5 < 36$, comparing against the node interval's lower bound, and $5 < 55$, comparing against the node's maximum value.
We recurse again on the left subtree from [29--29], as $5 < 29$ and $5 < 34$.
At the node [10--12] we find a match as $10 < 12$ and $12 < 10$.

\noindent
The complexity of this resolution algorithm is:
$$O(q(\log(n)+m))$$
where $q$ is the number of intervals in the query, $n$ is the number of intervals in the tree, and $m$ is the number of overlapping intervals with the query area (the size of the \texttt{address} set).
From this complexity, we can see why it's important that the space filling curve we use preservers locality and minimises the number of intervals we have to deal with.

To consider an extreme example of a large number of intervals, consider querying the entirety of a large music venue with many ubicomp devices.
This could overload the client, or the network, with too many results.
In practice, it may be prudent to have a maximum amount of addresses to be returned in a response, or limit the area clients can query, and instead return an error to the client prompting them to query with a more precise area.

Another scaling issue may be with the size of queries.
Ethernet's maximum transmission unit (MTU) of 1500 bytes is the de-facto standard in the Internet.
We have intentionally not specified an on the wire format for our queries, as this is left to future work, but 1500 bytes is only enough for 46 32-bit integers.
It is very possible that with our queries containing a list of all the Hilbert curve indices this may be too large to transmit in a single packet.
One solution to this would be to use TCP as a transport protocol as it doesn't have such size limitations.
Another would be to encode the geometry of the query area in a higher-level description than the Hilbert curve indices, such as a sphere with a radius at a particular coordinate, or a polygon made up of a number of coordinates, which could then be decoded into Hilbert curve indices when received.

We've described the spatial resolution mechanism of the SNS, but we haven't described what device will be receiving and responding to this query.
This shall be the topic of our next section.







%




\section{Spatial Architecture}
\label{sec-spatial-arch}

We have described a spatial mapping for resolving locations to network addresses algorithmically~(\S\ref{fig-spatial-map}), but this is not enough to complete the Spatial Name System (SNS).
We also need an associated system architecture for the SNS.
To draw an analogy, our spatial mapping is like the mechanism by which records are associated with a domain in DNS, but the architecture of the system is how zones are delegated, how recursive and iterative resolvers work, and how clients query the system.

\begin{figure}[p]
	\centering
	\includegraphics[width=0.75\textwidth]{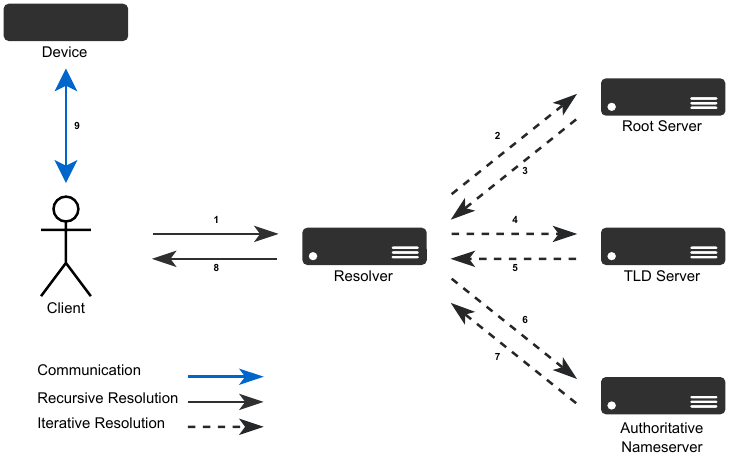}
	\caption[]{
		An example of the requests associated with a DNS lookup for a local device.
		The client requests the address of the device, through a recursive request to a DNS resolver.
		The DNS resolver makes iterate request to the root server, TLD server, and finally the authoritative nameserver.
		The result is returned to the client who can then communicate with the device.
	}
	\label{fig-dns}
\end{figure}

However, the existing DNS is inadequate for our AR interface into ubicomp.
The DNS has a strict hierarchical structure with centralized control, and authority is delegated based on zones.
Figure~\ref{fig-dns} shows an example of a DNS lookup for a local device which relies on this global infrastructure.
This is not an appropriate naming mechanism due to reliability, latency, security, and privacy issues~(\S\ref{sec-dns}).

\begin{figure}[p]
	\centering
	\includegraphics[width=0.75\textwidth]{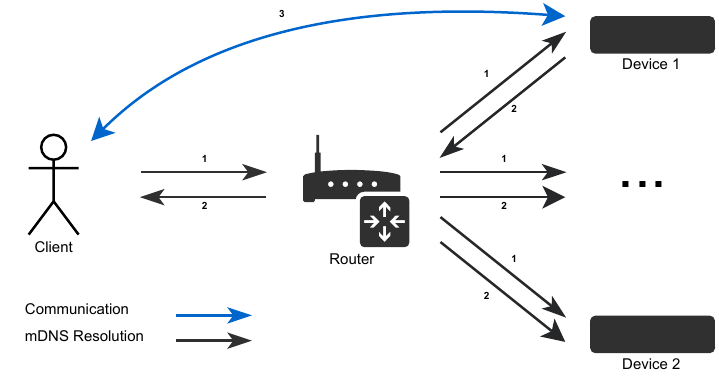}
	\caption[]{
		An example of the requests associated with an mDNS lookup for a local device.
		The client requests the address of Device 1, by sending a multicast message on the local network.
		The response is returned to the client and also multicasted on the network.
		All other devices participating in mDNS snoop on these packets.
	}
	\label{fig-mdns}
\end{figure}

\begin{figure}[t]
	\centering
	\includegraphics[width=0.5\textwidth]{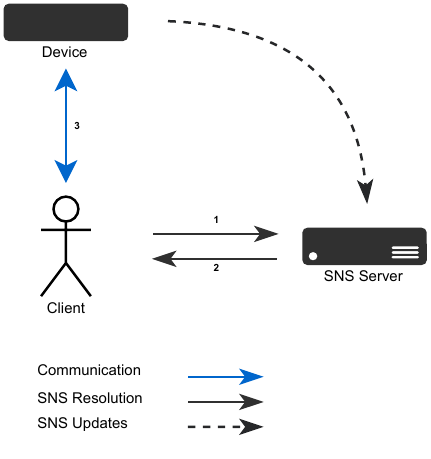}
	\caption[]{
		An example of the requests associated with a Spatial Name System (SNS) lookup for a local device.
		The client sends a request to the SNS server,
		the server uses the spatial mapping algorithm~(\S\ref{fig-spatial-map}) to map this to a network address,
		which is returned to the client.
		The client can then communicate with the device.
		Note that the device can update the server with its location directly and locally.
	}
	\label{fig-sns}
\end{figure}

In section~\ref{fig-spatial-map}, based on our experience with the LOC RR, we considered mDNS for our spatial mapping.
Mutlicast DNS is a decentralised protocol that resolves hostnames to IP addresses in a local network through UDP multicast~(\S\ref{fig-dns}).
Figure~\ref{fig-mdns} shows an example of an mDNS lookup for a local device.
It is well suited for service discovery in a local network, but this model of communication is ill-suited to our use case.
Consider the case from section~\ref{fig-spatial-map} where we are in a large music venue with many ubicomp devices.
There would be no mechanism for limiting the results returned which could flood the client or receiver with responses.
We are also constrained to a UDP broadcast domain, which may be an issue with devices using a disparate range of connectivity.
Finally, there are insufficient security and privacy considerations.

Instead, we propose a decentralised naming architecture that still uses a DNS server.
Figure~\ref{fig-sns} shows an example of a lookup in this system.
In section~\ref{sec-spatial-resolution} we assumed a coherent cache to map Hilbert curve index ranges to addresses.
The SNS server will maintain this augmented interval tree data structure.
The Client could find the SNS server through a fixed IP address, an extension to DHCP, or even using mDNS to bootstrap the process.

This will require support from the environment.
The server could be a dedicated device, a ubicomp device that is consistently available, or even could be added to a WiFi access point/routing device as increasing functionality is being pushed into the network.

This local DNS server could run a version of \texttt{ocaml-dns} modified with our spatial mapping algorithm, and an appropriate new resource record.
This would be suitable to run as a unikernel, especially if the device is an embedded device, for the benefits in performance and security \S(\ref{sec-technologies}).










\subsection{Scaling}
\label{sec-scaling}

While Hilbert curves are suitable for spatial mapping in a local environment, they don't address how to scale this architecture to naming across larger distances.
For example, while our spatial mapping algorithm is an appropriate mechanism for resolving locations in a room, we will run into complexity issues for a large number of devices due to our logarithmic complexity with respect to the number of intervals of address areas~(\S\ref{sec-resolution-mechanism})
For a large area, like a city, even with a low number of devices with a big location accuracy this could get computationally expensive quickly.

The environments we are considering for deploying our SNS are naturally segregated by physical boundaries, such as rooms, buildings, and cities.
We propose siloing the SNS into discrete `cells' corresponding to these physical domains.
We could use the existing DNS as a bridge between our SNS servers, as we are unlikely to require low latency communication with remote devices, but it may still be useful to name devices by their location; for example, a person querying what devices are in their office from home.
These multiple SNS instances would be interconnected through normal Internet routing.









\subsection{Mobility}
\label{sec-mobility}

We referred to a 3-dimensional mapping to a 1-dimensional network address when discussing the spatial mapping mechanism~(\S\ref{sec-spatial-resolution}).
However, there is a 4th dimension that we have touched on: time.
As ubicomp devices can be physically small and mobile, we need to consider how our architecture will support devices whose location changes over time.
First of all, we will need infrastructure to support positing systems for locating devices indoors with sufficient accuracy~(\S\ref{sec-local-positioning-systems}).
When a device moves location, it will update the SNS server with it's location directly as described in figure~\ref{fig-sns}.
This is analogous to dynamic DNS (DDNS), except having the entire system locally will remove the propagation time of DNS TTL values.
Once the SNS server receives this value, the \texttt{update} function of the spatial mapping mechanism~(\S\ref{fig-spatial-map}) will be used to update the map.

\subsection{Security and Privacy}
\label{sec-security-and-privacy}


The DNS LOC record has the following security considerations made of it:
\begin{quotation}
	\itshape
	High-precision LOC RR information could be used to plan a penetration of physical security, leading to potential denial-of-machine attacks. To avoid any appearance of suggesting this method to potential attackers, we declined the opportunity to name this RR ``ICBM''.\par
	\hfill \normalfont RFC 1876\\\hfill `A Means for Expressing Location Information in the Domain Name System'~\cite{dickinsonMeansExpressingLocation1996}
\end{quotation}
This highlights that location can be very sensitive data.
There are a few other notable issues with the DNS LOC: it is publicly available to anyone, and there is no verification that the location associated with a domain is accurate.

As we are discussing the security of the SNS, it is essential to define a threat model to understand the risks to our new architecture.
For example, using SNS will not prevent a malicious actor from physically accessing a device to compromise it.
The main attack vector for IoT devices is through remote access.
Physical attacks are possible, but our physical environments are already designed with security in mind, and these attacks do not scale.
A remote attacker can target millions of machines and networks with ease, whereas a physical attack takes time, planning and is much riskier for the attacker.
We shall therefore consider our threat model as consisting of how remote attackers might compromise our system and we rely on physicality to enable security as much as possible.
This is at odds with the DNS security extensions DNSSEC, which relies on a highly centralized model.

Like many early Internet Protocols, DNS was not designed with any security or privacy measures in mind, as it was developed in a much more cooperative environment than the modern Internet.
If a man-in-the-middle, such as your ISP, were to intercept your DNS query then they could return any desired IP address, such as a link to a phishing site.
While the use of higher-layer security mechanisms like TLS and HTTPS mitigate this, they are still vulnerable to Certificate Authority compromise.

\begin{figure}[b!]
	\centering
	\includegraphics[width=0.75\textwidth]{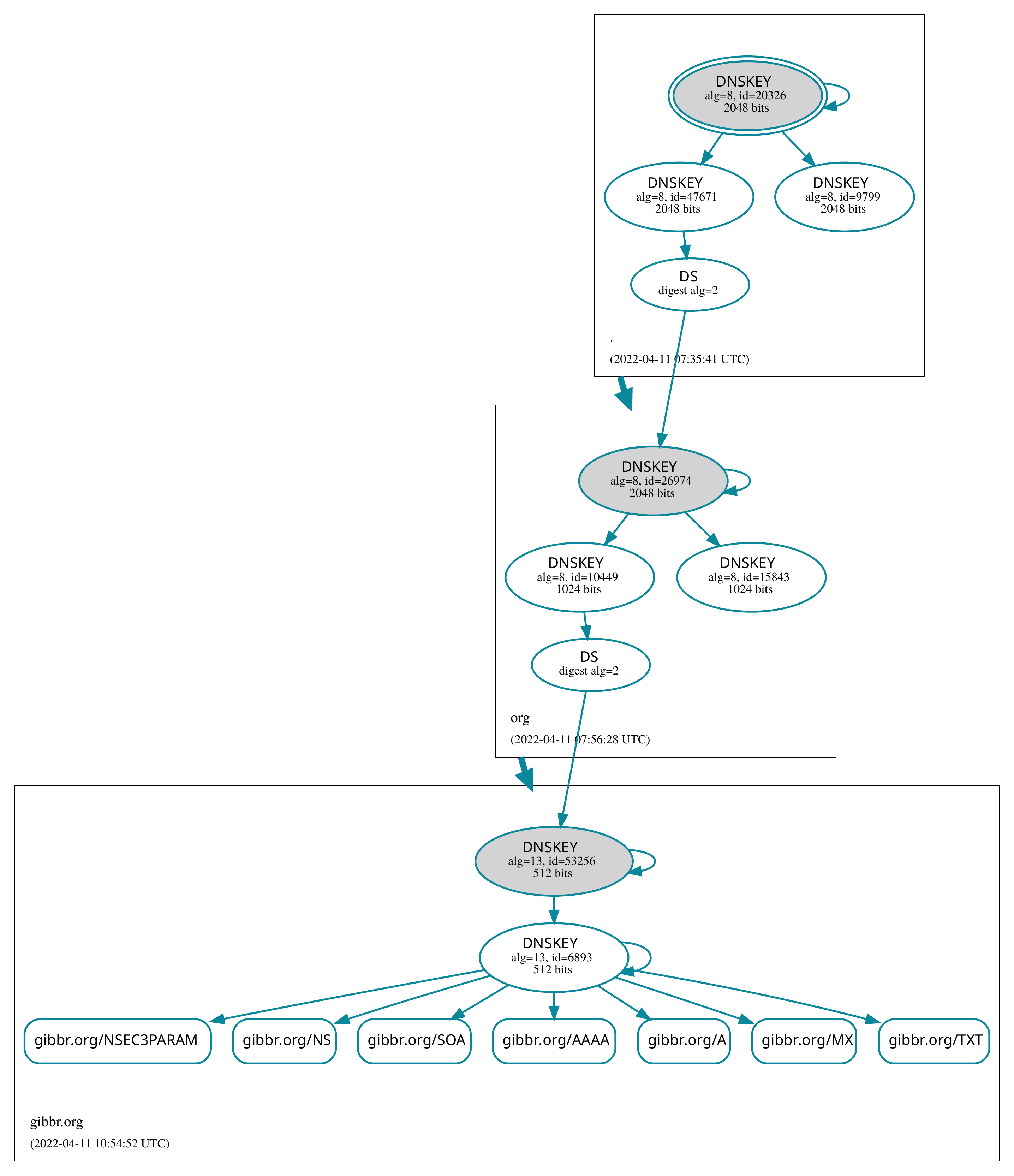}
	\caption{Visualisation of the DNSSEC status for the \texttt{gibbr.org} domain~\cite{GibbrOrgDNSViz}}
	\label{fig-dnssec}
\end{figure}

DNSSEC provides a set of security extensions to DNS to authenticate responses, so a resolver knows that the response was signed by the holder of a key signing key (KSK).
There is a hierarchical authentication model with each zone relying on a it's parent zone to server (and sign) a hash of it's zone signing key (ZSK) as a delegation signer (DS) record~\cite{roseDNSSecurityIntroduction2005}.
Note the ZSK is used to sign a KSK, so that the KSK can be frequently changed to meet security requirements, but the ZSK doesn't need to be updated with downtimes in DNSSEC due to DNS TTLs and propagating the change in DS record.

The root domain is signed in a root signing ceremony, where 7 people from ICANN and other trusted Internet community members sign the root zone's public keying information.
This ceremony is the source of trust for the entire system.
Figure~\ref{fig-dnssec} shows how this trust is propagated down the delegation tree for the domain \texttt{gibbr.org}.


As we want to enable local, optionally offline, operation, we don't want to rely on an external centralized source of trust like DNSSEC.
However, as long as we trust the local environment, we don't require authentication of the responses.
We are not querying across an Internet filled with censoring ISPs and snooping middleboxes.
This trust is informed by our threat model: we rely on the local physical environment to be trustworthy.

Additionally, DNSSEC's model of authentication would only authenticate messages from the SNS server in our case - but not the location updates from devices themselves.
A device could lie about where it is.
How could we stop a malicious or compromised local device from fabricating its position?
We could rely on the physicality offered by ubicomp.
One possibility is using ultrasound beacons broadcasting a one-time pad (OTP) from the SNS server, so the device has to physicality obtain this key from a room it's updating itself to be in~\cite{madhavapeddyContextAwareComputingSound2003a, madhavapeddyAudioNetworkingForgotten2005}.

Similarly, the same mechanism could be applied for access control on queries.
There could be a policy such that a person has to be physically present in a room to make a query of the devices there.
As mentioned with our threat model, attacks on this system would not scale due to its reliance on physical mechanisms.
It would also provide a much more seamless experience, more aligned with the vision of ubicomp~(\S\ref{sec-ubicomp}), if a user could simply authenticate themselves with their presence rather than solutions like clunky WiFi login portals that harvest user data.








Due to DNS's connectionless query mechanisms, using UDP, it's a prime target for Denial-of-Service (DoS) attack amplification.
This is a type of DoS attack where the attacker sends a request with a spoofed source IP to the server, which then responds to the spoofed IP address with a response larger than the original query.
Running many SNS servers that respond to Internet queries with large UDP responses could provide a whole new application vector for these attacks.
One limiting factor would be the relatively large SNS query sizes, compared to DNS queries, with the range of Hilbert curve indices~(\S\ref{fig-spatial-map}).
A larger query size reduces the application factor.
If the response is smaller than the query, this amplification factor will be less than 1.

Another mechanism to mitigate this could be authentication on queries.
A user would encrypt their authentication token with the SNS server's public key, and unauthenticated queries would be ignored.
This could be a standard username and password system, or even a token obtained from being in the location physicality within some timeframe.
This would only be useful for remote queries as due to the threat model we trust queries originating locally.

The final mitigation, possibly the most simple, would be to use a connection orientated transport layer protocol like TCP.
If the attacker uses a spoofed IP, the TCP 3-way handshake will fail.
The one packet sent to the spoofed IP - the SYN packet - will not result in any amplification.
Therefore, an application attack will not be possible with a connection orientated transport protocol.


While we have considered security in the context of authentication, authorisation and DoS mitigation, another very related issue is privacy.
Location can be a particularly sensitive type of information~\cite{weiserComputer21stCentury1999, beresfordLocationPrivacyUbiquitous}.
While the local positioning system~(\S\ref{sec-local-positioning-systems}) has its own privacy implications, transmitting this data over the network is what we're concerned with for the SNS.

Schemes like DNS over TLS (DoT), or DNS over HTTPS (DoH), encrypt DNS queries and prevent middlemen from snooping on queries.
We could use the same approach for SNS queries and updates.
It could still be possible for an actor to infer a client's location by snooping on the addresses they are communicating with.
Web proxies multiplex multiple hosts on one address to mitigate this, using TLS 1.3 encrypted Server Name Indicator (eSNI), or TLS encrypted ClientHello (ECH), to prevent the hostname being disclosed by the TLS handshake.
We could add a similar privacy-preserving proxy in our local network, but there would be a non-trivial complexity and overhead.

Having said all of this, many modern smartphones are live tracking users' locations across the world with GPS and storing this sensitive data in remote data centres~\cite{pressGoogleRecordsYour2018, naughtonForgetStateSurveillance2022}.
So even that only tracked your location locally would be a privacy improvement over this.
We should stress that there is no purely technical solution to privacy and any approaches must consider the social issues in their own right.

\section{Summary}
\label{sec-spatial-networking-summary}

In this chapter, we have explored the implementation of, and proposed the design of, the Spatial Name System.
We started by selecting suitable technologies to create the SNS~(\S\ref{sec-technologies}).
Then we implement the DNS LOC resource record in the OCaml DNS library~(\S\ref{sec-loc-rr}), starting with parsing the location information in sexagesimal latitude and longitude~(\S\ref{sec-parsing}), encoding the data in the RFC's specified on the wire format~(\S\ref{sec-encoding}), and thoroughly tested our implementation~(\S\ref{sec-testing}).
However, from implementing this record we determined that it was not suitable for the SNS, so we designed a spatial resolution system~(\S\ref{sec-spatial-resolution}) mapping higher-dimensional spaces to 1 dimension with Hilbert curves~(\S\ref{sec-spatial-mapping}) and resolving points on a Hilbert curve to network addresses with an augmented interval tree~(\S\ref{fig-spatial-resolution}).
We described the required architecture to make a workable system out of this algorithm~(\S\ref{sec-spatial-arch}), contrasting it with the existing DNS and mDNS systems, with special attention paid to scaling this system past local environments~(\S\ref{sec-scaling}), to support devices that move location~(\S\ref{sec-mobility}), and security and privacy considerations with this system~(\S\ref{sec-security-and-privacy}).

\chapter{Related Work}
\label{ch-related-work}

The vision of ubicomp has been an active area of research for decades.
However, we argue it has lost its way in recent years with the trend toward the Internet of Things, and we constrain ourselves to related work covered in the background~(\S\ref{sec-ubicomp}).
However, there has been other work related to this project that we shall summarise to put our work into context, highlight the novelty of the Spatial Name System, and provide further material for the curious reader.




The Main Name System (MNS)~\cite{deeganMainNameSystem2005} proposes an even more centralized DNS architecture, separating the administrative logic and distribution logic, for benefits in simplicity and removing inconsistencies
This work inspired the Spatial Name System in its radical rethinking of an archaic Internet protocol, but the SNS goes the opposite direction in proposing a radical \textit{decentralised} architecture, and does not maintain a consistent interface for clients and resolvers.

Another attempt at rethinking naming in the Internet includes the Intentional Naming System (INS)~\cite{adjie-winotoDesignImplementationIntentional1999}.
Published in 1999 --- the same year as `The Computer for the 21st Century'~\cite{weiserComputer21stCentury1999} --- this system proposes a radical rethinking of naming in the Internet with a decentralised system.
Similarly to the more modern and less revolutionary mDNS, it operates locally with decentralised nodes, and supports service discovery.
It also proposes an alternative naming mechanism to domain names, `Intentional names', which aim to address content by what the user is looking for ({\itshape `intent'}) rather than using hostnames or domain names.
These ideas developed into Information-Centric Networking.

The authors of the INS propose an example application `Floorplan', a service discovery tool showing how location-based services can be discovered using the INS.
An example request for this is:
\begin{lstlisting}
	[service=camera[entity=transmitter] [id=a]] [room=510]
\end{lstlisting}
\noindent
The proposed use case for this is a map displayed to the user with services represented with icons, which could be interpreted as prototypical of our augmented reality interface into ubicomp devices.

\clearpage
While our work is proposing decentralisation of naming in the Internet through local architecture, an alternative approach is using blockchain technologies.
The Ethereum Name Service (ENS)~\cite{NameResolverDistributed2021, xiaEthereumNameService2021} is one such approach.
The ENS maps human-readable names --- comparable to domain names --- to identifies like Ethereum addresses.
The \texttt{.eth} is `Ethereum native', but it also supports `importing' a DNS name into ENS for existing TLDs like \texttt{.com}.
This consists of proving ownership using DNSSEC and a TXT record to authenticate ownership of the domain.



The SNS's notion of resolving physical locations to network addresses is relatively novel, but work has been done on routing using physical location for MANETs with Greedy Perimeter Stateless Routing~\cite{karpGPSRGreedyPerimeter}.
This assumes a service for resolving network addresses to physical locations, which is the inverse of our required system.
Grid's location system~\cite{liScalableLocationService2000} is one such system that tracks mobile node locations in a distributed system.

Other work on geographic routing includes RFC 2009 `GPS-Based Addressing and Routing'~\cite{imielinskiGPSBasedAddressingRouting1996}.
This proposes routing mechanisms taking advantage of geographic information, such as multicasting to geographical regions and 
providing and advertising a service within a restricted area.
The encoding of areas consists of a series of points forming a polygon to perform `geographic addressing'.
The RFC proposes a mechanism for encoding geographical addressing, using the scheme:
\begin{lstlisting}
	site-code.city-code.state-code.country-code
\end{lstlisting}
\noindent


Finally, we shall mention the significant body of work on spatial indexing~\cite{moscoviterImprovingSpatialIndexing2016, azriReviewSpatialIndexing2013a, wardNewLocationTechnique1997}.
We have chosen a simple mechanism for our spatial mapping~(\S\ref{sec-spatial-resolution}) that is well suited to low latency updates and that can be updated incrementally, but it's very possible a better solution exists.




\chapter{Conclusions}
\label{ch-conclusions}


Recall from chapter~\ref{ch-introduction} that our vision with this project was an augmented reality (AR) interface into the world of ubicomp.
Our thesis statement was:
\begin{quotation}
	\itshape
	We have the hardware and software to support low latency augmented reality interactions, but the current network architecture is inadequate to support interconnecting them.
	We need a Spatial Name System that can map physical device locations to network addresses to overcome this limitation and unlock the potential of augmented reality.
\end{quotation}

Let us start by considering the first part of this statement that {\itshape `We have the hardware and software to support low latency augmented reality interactions but the current network architecture is inadequate to support interconnecting them'}.
In our background work~(\S\ref{ch-background}), we described this hardware and software in the form of ubicomp and IoT devices~(\S\ref{sec-iot}), as well as the HoloLens 2 augmented reality headset~(\S\ref{sec-augmented-reality}).
We described why the existing infrastructure is inadequate~(\S\ref{sec-ubicomp}), describing address solution mechanisms~(\S\ref{sec-arp}), how the current naming system came to be~(\S\ref{sec-hosts.txt}), and why the DNS is insufficient~(\S\ref{sec-dns}).
In our work with the HoloLens~(\S\ref{ch-augmented-reality-interface}) we tested this thesis by selecting cutting edge hardware~(\S\ref{sec-hw-and-sw}) and building a low latency AR program~(\S\ref{sec-development-process}) with mixed reality controls~(\S\ref{sec-mixed-reality-control}).
While working with the Holens~(\S\ref{sec-holographic-terminal}) we realise that the APIs are just not there to support our vision of an Augmented Reality Interface~(\S\ref{sec-interactivity-evaluation}).

Through this work, we've justified that {\itshape `We need a Spatial Name System that can map physical device locations to network addresses to overcome this limitation and unlock the potential of augmented reality.'}
We explore the design of this Spatial Name System (SNS)~(\S\ref{ch-spatial-networking}), justifying the technologies we've used towards this goal~(\S\ref{sec-technologies}).
We implement the DNS LOC record in the OCaml DNS library but find it is not suitable for our use case~(\S\ref{sec-loc-rr}).
We instead propose a spatial resolution mechanism~(\S\ref{sec-spatial-resolution}) to map physical spaces to a 1 dimensional representation~(\S\ref{sec-spatial-mapping}) and resolve these to network addresses efficiently~(\S\ref{fig-spatial-resolution}).
We describe the decentralised architecture of this system with infrastructure support~(\S\ref{sec-spatial-arch}) with considerations for scalability~(\S\ref{sec-scaling}), mobility~(\S\ref{sec-mobility}), security and privacy~(\S\ref{sec-security-and-privacy}).



\section{Future Work}
\label{sec:future-work}

A reoccurring issue throughout the practical parts of this project was reproducibility.
\begin{enumerate*}[(1)]
	\item The Active BAT system~(\S\ref{sec-local-positioning-systems}) was attempted to be revived, and the hardware was all still installed, but the software was no longer supported or in common use.
 	\item While developing for the HoloLens, setting up the development environment, as well as building a project once it was set up, was a tedious process of reading textual instructions and using performing imperative operations through GUIs~(\S\ref{sec-development-process}).
 	Additionally, regarding our experiment with Microsoft's environment, it was not possible to build the Windows Terminal locally for indeterminate reasons~(\S\ref{sec-holographic-terminal}).
	\item OCaml's package manager \texttt{opam} and build system \texttt{dune} were pleasant to use once one gained familiarity with them~(\S\ref{sec-loc-rr}), however, they could still benefit from having more reproducible environments.
\end{enumerate*}

We want to be able to rely on our infrastructure for not just 2 years, 3 years, or 4 years; but 20 years, 30 years, 40 years, or more.
Hardware obsoletion is not the problem it used to be, with the power of modern small energy-efficient computers, and as we are not seeing such drastic increases in computing performance as we used to --- Moore's law is declining.
Despite the obvious environmental and sustainability issues, having impermanent technological infrastructure has economic costs too.
To remedy this, the reproducible and declarative package and configuration manager NixOS~\cite{dolstraNixOSPurelyFunctional2010} has been used during this project.
The DNSSEC visualisation in figure~\ref{fig-dnssec} was obtained from an authoritative DNS nameserver running on NixOS (with Bind9, as Ocaml DNS doesn't support server-side DNSSEC).
Future work includes exploring how this can benefit our Spatial Name System and ubicomp as a whole.



Other future work includes exploring the HoloLens spatial mapping APIs in more depth including object tracking, object placing and navigation.
In this project, once we had a basic game working, we focused on trying to network the device.

Another area is exploring the sophisticated spatial indexing techniques mentioned in chapter~\ref{ch-related-work}.
We choose to design a mechanism for our spatial resolution which was well suited to low latency efficient lookups~(\S\ref{sec-spatial-resolution}), but there is certainly the possibility of finding a more appropriate algorithm.

And finally, the most obvious aspect for future work is to implement and experiment with the Spatial Name System.
Unfortunately due to the time constraints of the project we were unable to complete this.
To this end, I'm planning on continuing this area of research in my doctoral studies.

\label{lastpage}


\printbibliography[title={References}]

@inproceedings{adjie-winotoDesignImplementationIntentional1999,
  title = {The Design and Implementation of an Intentional Naming System},
  booktitle = {Proceedings of the Seventeenth {{ACM}} Symposium on {{Operating}} Systems Principles},
  author = {Adjie-Winoto, William and Schwartz, Elliot and Balakrishnan, Hari and Lilley, Jeremy},
  date = {1999-12-12},
  series = {{{SOSP}} '99},
  pages = {186--201},
  publisher = {{Association for Computing Machinery}},
  location = {{New York, NY, USA}},
  doi = {10.1145/319151.319164},
  url = {https://doi.org/10.1145/319151.319164},
  urldate = {2022-06-05},
  isbn = {978-1-58113-140-6}
}

@inproceedings{affeldtFormalNetworkPacket2012,
  title = {Formal Network Packet Processing with Minimal Fuss: Invertible Syntax Descriptions at Work},
  shorttitle = {Formal Network Packet Processing with Minimal Fuss},
  booktitle = {Proceedings of the Sixth Workshop on {{Programming}} Languages Meets Program Verification - {{PLPV}} '12},
  author = {Affeldt, Reynald and Nowak, David and Oiwa, Yutaka},
  date = {2012},
  pages = {27},
  publisher = {{ACM Press}},
  location = {{Philadelphia, Pennsylvania, USA}},
  doi = {10.1145/2103776.2103781},
  url = {http://dl.acm.org/citation.cfm?doid=2103776.2103781},
  urldate = {2022-06-02},
  eventtitle = {The Sixth Workshop},
  isbn = {978-1-4503-1125-0},
  langid = {english}
}

@book{azriReviewSpatialIndexing2013a,
  title = {Review of {{Spatial Indexing Techniques}} for {{Large Urban Data Management}}},
  author = {Azri, Suhaibah and Ujang, Uznir and Anton, François and Mioc, Darka and Rahman, Alias},
  date = {2013-01-01}
}

@article{backusRevisedReportAlgorithm1963,
  title = {Revised Report on the Algorithm Language {{ALGOL}} 60},
  author = {Backus, J. W. and Bauer, F. L. and Green, J. and Katz, C. and McCarthy, J. and Perlis, A. J. and Rutishauser, H. and Samelson, K. and Vauquois, B. and Wegstein, J. H. and van Wijngaarden, A. and Woodger, M.},
  editor = {Naur, P.},
  options = {useprefix=true},
  date = {1963-01},
  journaltitle = {Communications of the ACM},
  shortjournal = {Commun. ACM},
  volume = {6},
  number = {1},
  pages = {1--17},
  issn = {0001-0782, 1557-7317},
  doi = {10.1145/366193.366201},
  url = {https://dl.acm.org/doi/10.1145/366193.366201},
  urldate = {2022-06-02},
  langid = {english}
}

@inproceedings{barhamXenArtVirtualization2003,
  title = {Xen and the Art of Virtualization},
  booktitle = {Proceedings of the Nineteenth {{ACM}} Symposium on {{Operating}} Systems Principles},
  author = {Barham, Paul and Dragovic, Boris and Fraser, Keir and Hand, Steven and Harris, Tim and Ho, Alex and Neugebauer, Rolf and Pratt, Ian and Warfield, Andrew},
  date = {2003-10-19},
  series = {{{SOSP}} '03},
  pages = {164--177},
  publisher = {{Association for Computing Machinery}},
  location = {{New York, NY, USA}},
  doi = {10.1145/945445.945462},
  url = {https://doi.org/10.1145/945445.945462},
  urldate = {2022-01-25},
  isbn = {978-1-58113-757-6}
}

@thesis{beresfordLocationPrivacyUbiquitous,
  title = {Location Privacy in Ubiquitous Computing},
  author = {Beresford, Alastair R},
  langid = {english}
}

@artwork{boloEnglishIBMPersonal2014,
  title = {English:  {{IBM Personal Computer}}},
  shorttitle = {English},
  author = {Bolo, Rama {and} Musée},
  date = {2014-11-13},
  url = {https://commons.wikimedia.org/wiki/File:IBM_PC-IMG_7271_(transparent).png},
  urldate = {2022-05-21}
}

@report{bradenRequirementsInternetHosts1989,
  type = {Request for Comments},
  title = {Requirements for {{Internet Hosts}} - {{Application}} and {{Support}}},
  author = {Braden, Robert T.},
  date = {1989-10},
  number = {RFC 1123},
  institution = {{Internet Engineering Task Force}},
  doi = {10.17487/RFC1123},
  url = {https://datatracker.ietf.org/doc/rfc1123},
  urldate = {2022-06-02},
  pagetotal = {98}
}

@report{cheshireMulticastDNS2013,
  type = {Request for Comments},
  title = {Multicast {{DNS}}},
  author = {Cheshire, Stuart and Krochmal, Marc},
  date = {2013-02},
  number = {RFC 6762},
  institution = {{Internet Engineering Task Force}},
  doi = {10.17487/RFC6762},
  url = {https://datatracker.ietf.org/doc/rfc6762},
  urldate = {2022-05-12},
  pagetotal = {70}
}

@article{coleHalftoningDitherEdge2005,
  title = {Halftoning without Dither or Edge Enhancement},
  author = {Cole, A. J.},
  date = {2005},
  journaltitle = {The Visual Computer},
  doi = {10.1007/BF01905689}
}

@online{consortiumBIND,
  title = {{{BIND}} 9},
  author = {Consortium, Internet Systems},
  url = {https://www.isc.org/bind/},
  urldate = {2022-06-01},
  langid = {american}
}

@inproceedings{deberHowMuchFaster2015,
  title = {How {{Much Faster}} Is {{Fast Enough}}?: {{User Perception}} of {{Latency}} \& {{Latency Improvements}} in {{Direct}} and {{Indirect Touch}}},
  shorttitle = {How {{Much Faster}} Is {{Fast Enough}}?},
  booktitle = {Proceedings of the 33rd {{Annual ACM Conference}} on {{Human Factors}} in {{Computing Systems}}},
  author = {Deber, Jonathan and Jota, Ricardo and Forlines, Clifton and Wigdor, Daniel},
  date = {2015-04-18},
  pages = {1827--1836},
  publisher = {{ACM}},
  location = {{Seoul Republic of Korea}},
  doi = {10.1145/2702123.2702300},
  url = {https://dl.acm.org/doi/10.1145/2702123.2702300},
  urldate = {2021-12-31},
  eventtitle = {{{CHI}} '15: {{CHI Conference}} on {{Human Factors}} in {{Computing Systems}}},
  isbn = {978-1-4503-3145-6},
  langid = {english}
}

@article{deeganMainNameSystem2005,
  title = {The Main Name System: An Exercise in Centralized Computing},
  shorttitle = {The Main Name System},
  author = {Deegan, Tim and Crowcroft, Jon and Warfield, Andrew},
  date = {2005-10-06},
  journaltitle = {ACM SIGCOMM Computer Communication Review},
  shortjournal = {SIGCOMM Comput. Commun. Rev.},
  volume = {35},
  number = {5},
  pages = {5--14},
  issn = {0146-4833},
  doi = {10.1145/1096536.1096538},
  url = {https://dl.acm.org/doi/10.1145/1096536.1096538},
  urldate = {2021-12-08},
  langid = {english}
}

@report{dickinsonMeansExpressingLocation1996,
  type = {Request for Comments},
  title = {A {{Means}} for {{Expressing Location Information}} in the {{Domain Name System}}},
  author = {Dickinson, Ian and Vixie, Paul A. and Davis, Christopher and Goodwin, Tim},
  date = {1996-01},
  number = {RFC 1876},
  institution = {{Internet Engineering Task Force}},
  doi = {10.17487/RFC1876},
  url = {https://datatracker.ietf.org/doc/rfc1876},
  urldate = {2022-01-18},
  pagetotal = {18}
}

@thesis{dolstraNixOSPurelyFunctional2010,
  title = {{{NixOS}}: {{A Purely Functional Linux Distribution}}},
  author = {Dolstra, Eelco},
  date = {2010},
  langid = {english}
}

@report{DomainNamesConcepts1983,
  type = {Request for Comments},
  title = {Domain Names: {{Concepts}} and Facilities},
  shorttitle = {Domain Names},
  date = {1983-11},
  number = {RFC 882},
  institution = {{Internet Engineering Task Force}},
  doi = {10.17487/RFC0882},
  url = {https://datatracker.ietf.org/doc/rfc882},
  urldate = {2022-05-15},
  pagetotal = {31}
}

@report{DomainNamesImplementation1987,
  type = {Request for Comments},
  title = {Domain Names - Implementation and Specification},
  date = {1987-11},
  number = {RFC 1035},
  institution = {{Internet Engineering Task Force}},
  doi = {10.17487/RFC1035},
  url = {https://datatracker.ietf.org/doc/rfc1035},
  urldate = {2022-05-15},
  pagetotal = {55}
}

@online{ErrorWhenBuilding,
  title = {Error When Building · {{Issue}} \#12219 · Microsoft/Terminal},
  url = {https://github.com/microsoft/terminal/issues/12219},
  urldate = {2022-05-22},
  langid = {english},
  organization = {{GitHub}}
}

@online{GetMSIXPackaging,
  title = {Get \${{MSIX Packaging Tool}} from the {{Microsoft Store}}},
  url = {https://apps.microsoft.com/store/detail/msix-packaging-tool/9N5LW3JBCXKF?hl=en-gb&gl=GB},
  urldate = {2022-05-28},
  langid = {british}
}

@online{gibbCubes,
  title = {Cubes},
  author = {Gibb, Ryan},
  url = {https://ryan.freumh.org/cubes/},
  urldate = {2022-05-26}
}

@software{gibbCubesGithub2022,
  title = {Cubes {{Github}}},
  author = {Gibb, Ryan},
  date = {2022-01-13T19:52:30Z},
  origdate = {2021-05-25T01:48:13Z},
  url = {https://github.com/RyanGibb/cubes},
  urldate = {2022-05-15}
}

@online{GibbrOrgDNSViz,
  title = {Gibbr.Org | {{DNSViz}}},
  url = {https://dnsviz.net/d/gibbr.org/dnssec/},
  urldate = {2022-05-21}
}

@online{HackersHeroesComputer,
  title = {Hackers: Heroes of the Computer Revolution: | {{Guide}} Books},
  url = {https://dl.acm.org/doi/10.5555/1925},
  urldate = {2022-06-04}
}

@online{hamalawiDevicePortalAPI,
  title = {Device Portal {{API}} Reference - {{Mixed Reality}}},
  author = {{hamalawi}},
  url = {https://docs.microsoft.com/en-us/windows/mixed-reality/develop/advanced-concepts/device-portal-api-reference},
  urldate = {2022-05-28},
  langid = {american}
}

@article{harterAnatomyContextAwareApplication2002,
  title = {The {{Anatomy}} of a {{Context-Aware Application}}},
  author = {Harter, Andy and Hopper, Andy and Steggles, Pete and Ward, Andy and Webster, Paul},
  date = {2002-03-01},
  journaltitle = {Wireless Networks},
  shortjournal = {Wireless Networks},
  volume = {8},
  number = {2},
  pages = {187--197},
  issn = {1572-8196},
  doi = {10.1023/A:1013767926256},
  url = {https://doi.org/10.1023/A:1013767926256},
  urldate = {2022-05-20},
  langid = {english}
}

@article{hilbertUeberStetigeAbbildung1891,
  title = {Ueber die stetige Abbildung einer Line auf ein Flächenstück},
  author = {Hilbert, David},
  date = {1891-09-01},
  journaltitle = {Mathematische Annalen},
  shortjournal = {Math. Ann.},
  volume = {38},
  number = {3},
  pages = {459--460},
  issn = {1432-1807},
  doi = {10.1007/BF01199431},
  url = {https://doi.org/10.1007/BF01199431},
  urldate = {2022-06-02},
  langid = {german}
}

@online{HoloLens,
  title = {{{HoloLens}} 2},
  url = {https://www.microsoft.com/en-us/hololens/buy},
  urldate = {2022-05-21},
  langid = {american}
}

@online{HololensBuildIssue,
  title = {Hololens Build · {{Issue}} \#9267 · Microsoft/Terminal},
  url = {https://github.com/microsoft/terminal/issues/9267},
  urldate = {2022-05-28},
  langid = {english},
  organization = {{GitHub}}
}

@report{HostNamesOnline1973,
  type = {Request for Comments},
  title = {Host Names On-Line 1973},
  date = {1973-12},
  number = {RFC 606},
  institution = {{Internet Engineering Task Force}},
  doi = {10.17487/RFC0606},
  url = {https://datatracker.ietf.org/doc/rfc606},
  urldate = {2022-06-02},
  pagetotal = {3}
}

@report{HostNamesOnline1974,
  type = {Request for Comments},
  title = {Host Names On-Line},
  date = {1974-01},
  number = {RFC 608},
  institution = {{Internet Engineering Task Force}},
  doi = {10.17487/RFC0608},
  url = {https://datatracker.ietf.org/doc/rfc608},
  urldate = {2022-05-11},
  pagetotal = {4}
}

@report{imielinskiGPSBasedAddressingRouting1996,
  type = {Request for Comments},
  title = {{{GPS-Based Addressing}} and {{Routing}}},
  author = {Imielinski, Tomaz and Navas, Julio C.},
  date = {1996-11},
  number = {RFC 2009},
  institution = {{Internet Engineering Task Force}},
  doi = {10.17487/RFC2009},
  url = {https://datatracker.ietf.org/doc/rfc2009},
  urldate = {2022-05-15},
  pagetotal = {27}
}

@article{karpGPSRGreedyPerimeter,
  title = {{{GPSR}}: {{Greedy Perimeter Stateless Routing}} for {{Wireless Networks}}},
  author = {Karp, Brad and Kung, H T},
  pages = {12},
  langid = {english}
}

@report{lawrenceObsoletingIQUERY2002,
  type = {Request for Comments},
  title = {Obsoleting {{IQUERY}}},
  author = {Lawrence, David C.},
  date = {2002-11},
  number = {RFC 3425},
  institution = {{Internet Engineering Task Force}},
  doi = {10.17487/RFC3425},
  url = {https://datatracker.ietf.org/doc/rfc3425},
  urldate = {2022-06-02},
  pagetotal = {5}
}

@inproceedings{liScalableLocationService2000,
  title = {A Scalable Location Service for Geographic Ad Hoc Routing},
  booktitle = {Proceedings of the 6th Annual International Conference on {{Mobile}} Computing and Networking  - {{MobiCom}} '00},
  author = {Li, Jinyang and Jannotti, John and De Couto, Douglas S. J. and Karger, David R. and Morris, Robert},
  date = {2000},
  pages = {120--130},
  publisher = {{ACM Press}},
  location = {{Boston, Massachusetts, United States}},
  doi = {10.1145/345910.345931},
  url = {http://portal.acm.org/citation.cfm?doid=345910.345931},
  urldate = {2021-12-28},
  eventtitle = {The 6th Annual International Conference},
  isbn = {978-1-58113-197-0},
  langid = {english}
}

@inproceedings{madhavapeddyArchitectureInterspatialCommunication2018,
  title = {An Architecture for Interspatial Communication},
  booktitle = {{{IEEE INFOCOM}} 2018 - {{IEEE Conference}} on {{Computer Communications Workshops}} ({{INFOCOM WKSHPS}})},
  author = {Madhavapeddy, Anil and Sivaramakrishnan, K C and Gordon, Gemma and Gazagnaire, Thomas},
  date = {2018-04},
  pages = {716--723},
  publisher = {{IEEE}},
  location = {{Honolulu, HI}},
  doi = {10.1109/INFCOMW.2018.8406931},
  url = {https://ieeexplore.ieee.org/document/8406931/},
  urldate = {2021-11-02},
  eventtitle = {2018 {{IEEE Conference}} on {{Computer Communications Workshops}} ({{INFOCOM WKSHPS}})},
  isbn = {978-1-5386-5979-3},
  langid = {english}
}

@article{madhavapeddyAudioNetworkingForgotten2005,
  title = {Audio Networking: The Forgotten Wireless Technology},
  shorttitle = {Audio Networking},
  author = {Madhavapeddy, A. and Sharp, R. and Scott, D. and Tse, A.},
  date = {2005-07},
  journaltitle = {IEEE Pervasive Computing},
  volume = {4},
  number = {3},
  pages = {55--60},
  issn = {1558-2590},
  doi = {10.1109/MPRV.2005.50},
  eventtitle = {{{IEEE Pervasive Computing}}}
}

@incollection{madhavapeddyContextAwareComputingSound2003a,
  title = {Context-{{Aware Computing}} with {{Sound}}},
  booktitle = {{{UbiComp}} 2003: {{Ubiquitous Computing}}},
  author = {Madhavapeddy, Anil and Scott, David and Sharp, Richard},
  editor = {Dey, Anind K. and Schmidt, Albrecht and McCarthy, Joseph F.},
  options = {useprefix=true},
  date = {2003},
  series = {Lecture {{Notes}} in {{Computer Science}}},
  volume = {2864},
  pages = {315--332},
  publisher = {{Springer Berlin Heidelberg}},
  location = {{Berlin, Heidelberg}},
  doi = {10.1007/978-3-540-39653-6_25},
  url = {http://link.springer.com/10.1007/978-3-540-39653-6_25},
  urldate = {2022-02-06},
  editorb = {Goos, Gerhard and Hartmanis, Juris and van Leeuwen, Jan},
  editorbtype = {redactor},
  isbn = {978-3-540-20301-8 978-3-540-39653-6},
  langid = {english}
}

@article{madhavapeddyMelangeCreatingFunctional2007,
  title = {Melange: Creating a "Functional" Internet},
  shorttitle = {Melange},
  author = {Madhavapeddy, Anil and Ho, Alex and Deegan, Tim and Scott, David and Sohan, Ripduman},
  date = {2007-03-21},
  journaltitle = {ACM SIGOPS Operating Systems Review},
  shortjournal = {SIGOPS Oper. Syst. Rev.},
  volume = {41},
  number = {3},
  pages = {101--114},
  issn = {0163-5980},
  doi = {10.1145/1272998.1273009},
  url = {https://doi.org/10.1145/1272998.1273009},
  urldate = {2022-02-10}
}

@incollection{madhavapeddyStudyBluetoothPropagation2005,
  title = {A {{Study}} of {{Bluetooth Propagation Using Accurate Indoor Location Mapping}}},
  booktitle = {{{UbiComp}} 2005: {{Ubiquitous Computing}}},
  author = {Madhavapeddy, Anil and Tse, Alastair},
  editor = {Beigl, Michael and Intille, Stephen and Rekimoto, Jun and Tokuda, Hideyuki},
  date = {2005},
  series = {Lecture {{Notes}} in {{Computer Science}}},
  volume = {3660},
  pages = {105--122},
  publisher = {{Springer Berlin Heidelberg}},
  location = {{Berlin, Heidelberg}},
  doi = {10.1007/11551201_7},
  url = {http://link.springer.com/10.1007/11551201_7},
  urldate = {2022-05-25},
  editorb = {Hutchison, David and Kanade, Takeo and Kittler, Josef and Kleinberg, Jon M. and Mattern, Friedemann and Mitchell, John C. and Naor, Moni and Nierstrasz, Oscar and Pandu Rangan, C. and Steffen, Bernhard and Sudan, Madhu and Terzopoulos, Demetri and Tygar, Dough and Vardi, Moshe Y. and Weikum, Gerhard},
  editorbtype = {redactor},
  isbn = {978-3-540-28760-5 978-3-540-31941-2},
  langid = {english}
}

@article{madhavapeddyUnikernelsLibraryOperating2013,
  title = {Unikernels: Library Operating Systems for the Cloud},
  shorttitle = {Unikernels},
  author = {Madhavapeddy, Anil and Mortier, Richard and Rotsos, Charalampos and Scott, David and Singh, Balraj and Gazagnaire, Thomas and Smith, Steven and Hand, Steven and Crowcroft, Jon},
  date = {2013-03-16},
  journaltitle = {ACM SIGARCH Computer Architecture News},
  shortjournal = {SIGARCH Comput. Archit. News},
  volume = {41},
  number = {1},
  pages = {461--472},
  issn = {0163-5964},
  doi = {10.1145/2490301.2451167},
  url = {https://doi.org/10.1145/2490301.2451167},
  urldate = {2022-01-25}
}

@online{MigratingUNetNetcode,
  title = {Migrating {{From UNet}} to {{Netcode}} for {{GameObjects}} | {{Unity Multiplayer Networking}}},
  url = {https://docs-multiplayer.unity3d.com//netcode/current/migration/migratingtonetcode},
  urldate = {2022-05-29},
  langid = {english}
}

@article{milgramAugmentedRealityClass1994,
  title = {Augmented Reality: {{A}} Class of Displays on the Reality-Virtuality Continuum},
  shorttitle = {Augmented Reality},
  author = {Milgram, Paul and Takemura, Haruo and Utsumi, Akira and Kishino, Fumio},
  date = {1994-01-01},
  journaltitle = {Telemanipulator and Telepresence Technologies},
  shortjournal = {Telemanipulator and Telepresence Technologies},
  volume = {2351},
  doi = {10.1117/12.197321}
}

@incollection{moscoviterImprovingSpatialIndexing2016,
  title = {Improving {{Spatial Indexing}} and {{Searching}} for {{Location-Based DNS Queries}}},
  booktitle = {Wired/{{Wireless Internet Communications}}},
  author = {Moscoviter, Daniel and Gholibeigi, Mozhdeh and Meijerink, Bernd and Kooijman, Ruben and Krijger, Paul and Heijenk, Geert},
  editor = {Mamatas, Lefteris and Matta, Ibrahim and Papadimitriou, Panagiotis and Koucheryavy, Yevgeni},
  date = {2016},
  series = {Lecture {{Notes}} in {{Computer Science}}},
  volume = {9674},
  pages = {187--198},
  publisher = {{Springer International Publishing}},
  location = {{Cham}},
  doi = {10.1007/978-3-319-33936-8_15},
  url = {http://link.springer.com/10.1007/978-3-319-33936-8_15},
  urldate = {2022-05-13},
  isbn = {978-3-319-33935-1 978-3-319-33936-8},
  langid = {english}
}

@online{NameResolverDistributed2021,
  title = {A {{Name Resolver}} for the {{Distributed Web}}},
  date = {2021-01-13T12:00:00},
  url = {http://blog.cloudflare.com/cloudflare-distributed-web-resolver/},
  urldate = {2022-05-12},
  langid = {english},
  organization = {{The Cloudflare Blog}}
}

@article{naughtonForgetStateSurveillance2022,
  title = {Forget State Surveillance. {{Our}} Tracking Devices Are Now Doing the Same Job},
  author = {Naughton, John},
  date = {2022-02-19T16:00:38},
  journaltitle = {The Observer},
  issn = {0029-7712},
  url = {https://www.theguardian.com/commentisfree/2022/feb/19/forget-state-surveillance-our-tracking-devices-are-now-doing-the-same-job},
  urldate = {2022-06-05},
  entrysubtype = {newspaper},
  journalsubtitle = {Opinion},
  langid = {british}
}

@online{OpamDns,
  title = {Opam - Dns},
  url = {https://opam.ocaml.org/packages/dns/},
  urldate = {2022-06-01}
}

@article{peanoCourbeQuiRemplit1890,
  title = {Sur une courbe, qui remplit toute une aire plane},
  author = {Peano, G.},
  date = {1890-03-01},
  journaltitle = {Mathematische Annalen},
  shortjournal = {Math. Ann.},
  volume = {36},
  number = {1},
  pages = {157--160},
  issn = {1432-1807},
  doi = {10.1007/BF01199438},
  url = {https://doi.org/10.1007/BF01199438},
  urldate = {2022-06-03},
  langid = {french}
}

@article{pressGoogleRecordsYour2018,
  title = {Google Records Your Location Even When You Tell It Not To},
  author = {Press, Associated},
  date = {2018-08-13T18:30:19},
  journaltitle = {The Guardian},
  issn = {0261-3077},
  url = {https://www.theguardian.com/technology/2018/aug/13/google-location-tracking-android-iphone-mobile},
  urldate = {2022-06-05},
  entrysubtype = {newspaper},
  journalsubtitle = {Technology},
  langid = {british}
}

@inproceedings{priyanthaCricketLocationsupportSystem2000,
  title = {The {{Cricket}} Location-Support System},
  booktitle = {Proceedings of the 6th Annual International Conference on {{Mobile}} Computing and Networking  - {{MobiCom}} '00},
  author = {Priyantha, Nissanka B. and Chakraborty, Anit and Balakrishnan, Hari},
  date = {2000},
  pages = {32--43},
  publisher = {{ACM Press}},
  location = {{Boston, Massachusetts, United States}},
  doi = {10.1145/345910.345917},
  url = {http://portal.acm.org/citation.cfm?doid=345910.345917},
  urldate = {2021-12-27},
  eventtitle = {The 6th Annual International Conference},
  isbn = {978-1-58113-197-0},
  langid = {english}
}

@online{QuestOurMost,
  title = {Quest 2: {{Our}} Most Advanced New All-in-One {{VR}} Headset | {{Meta Quest}}},
  shorttitle = {Quest 2},
  url = {https://store.facebook.com/gb/quest/products/quest-2/},
  urldate = {2022-05-21},
  langid = {english}
}

@article{rehmanInterfacingInvisibleComputer2002,
  title = {Interfacing with the {{Invisible Computer}}},
  author = {Rehman, Kasim and Stajano, Frank and Coulouris, George},
  date = {2002},
  pages = {4},
  langid = {english}
}

@report{roseDNSSecurityIntroduction2005,
  type = {Request for Comments},
  title = {{{DNS Security Introduction}} and {{Requirements}}},
  author = {Rose, Scott and Larson, Matt and Massey, Dan and Austein, Rob and Arends, Roy},
  date = {2005-03},
  number = {RFC 4033},
  institution = {{Internet Engineering Task Force}},
  doi = {10.17487/RFC4033},
  url = {https://datatracker.ietf.org/doc/rfc4033},
  urldate = {2022-05-19},
  pagetotal = {21}
}

@report{schulzeDNSEncodingGeographical1994,
  type = {Request for Comments},
  title = {{{DNS Encoding}} of {{Geographical Location}}},
  author = {Schulze, Mike and Farrell, Craig and Baldoni, Daniel and Pleitner, Scott},
  date = {1994-11},
  number = {RFC 1712},
  institution = {{Internet Engineering Task Force}},
  doi = {10.17487/RFC1712},
  url = {https://datatracker.ietf.org/doc/rfc1712},
  urldate = {2022-05-13},
  pagetotal = {7}
}

@article{scottUsingVisualTags2005,
  title = {Using Visual Tags to Bypass {{Bluetooth}} Device Discovery},
  author = {Scott, David and Sharp, Richard and Madhavapeddy, Anil and Upton, Eben},
  date = {2005-01},
  journaltitle = {ACM SIGMOBILE Mobile Computing and Communications Review},
  shortjournal = {SIGMOBILE Mob. Comput. Commun. Rev.},
  volume = {9},
  number = {1},
  pages = {41--53},
  issn = {1559-1662, 1931-1222},
  doi = {10.1145/1055959.1055965},
  url = {https://dl.acm.org/doi/10.1145/1055959.1055965},
  urldate = {2022-05-14},
  langid = {english}
}

@misc{sissonDnsSurvey20102010,
  title = {Dns\_survey\_2010.Pdf},
  author = {Sisson, Geoffrey},
  date = {2010},
  publisher = {{Tech. rep., The Measurement Factor}},
  url = {http://www.open-spf.org/surveys/201010/dns_survey_2010.pdf},
  urldate = {2022-05-31}
}

@book{stephensonSnowCrash1992,
  title = {Snow {{Crash}}},
  author = {Stephenson, Neal},
  date = {1992-06},
  publisher = {{Bantam Books}}
}

@online{stevewhimsWindowsNetworkingServiceDiscovery,
  title = {Windows.{{Networking}}.{{ServiceDiscovery}}.{{Dnssd Namespace}} - {{Windows UWP}} Applications},
  author = {{stevewhims}},
  url = {https://docs.microsoft.com/en-us/uwp/api/windows.networking.servicediscovery.dnssd},
  urldate = {2022-05-29},
  langid = {american}
}

@article{tariqNonGPSPositioningSystems2017,
  title = {Non-{{GPS Positioning Systems}}: {{A Survey}}},
  shorttitle = {Non-{{GPS Positioning Systems}}},
  author = {Tariq, Zain Bin and Cheema, Dost Muhammad and Kamran, Muhammad Zahir and Naqvi, Ijaz Haider},
  date = {2017-11-08},
  journaltitle = {ACM Computing Surveys},
  shortjournal = {ACM Comput. Surv.},
  volume = {50},
  number = {4},
  pages = {1--34},
  issn = {0360-0300, 1557-7341},
  doi = {10.1145/3098207},
  url = {https://dl.acm.org/doi/10.1145/3098207},
  urldate = {2021-12-31},
  langid = {english}
}

@article{taylorFacebookOutageWhat2021,
  title = {Facebook Outage: What Went Wrong and Why Did It Take so Long to Fix after Social Platform Went Down?},
  shorttitle = {Facebook Outage},
  author = {Taylor, Josh},
  date = {2021-10-05T05:53:53},
  journaltitle = {The Guardian},
  issn = {0261-3077},
  url = {https://www.theguardian.com/technology/2021/oct/05/facebook-outage-what-went-wrong-and-why-did-it-take-so-long-to-fix},
  urldate = {2022-05-18},
  entrysubtype = {newspaper},
  journalsubtitle = {Technology},
  langid = {british}
}

@online{thetuvixInstallToolsMixed,
  title = {Install the Tools - {{Mixed Reality}}},
  author = {{thetuvix}},
  url = {https://docs.microsoft.com/en-us/windows/mixed-reality/develop/install-the-tools},
  urldate = {2022-05-27},
  langid = {american}
}

@online{thetuvixNativeDevelopmentOverview,
  title = {Native Development Overview - {{Mixed Reality}}},
  author = {{thetuvix}},
  url = {https://docs.microsoft.com/en-us/windows/mixed-reality/develop/native/directx-development-overview},
  urldate = {2022-05-29},
  langid = {american}
}

@online{TransportsUnityMultiplayer,
  title = {Transports | {{Unity Multiplayer Networking}}},
  url = {https://docs-multiplayer.unity3d.com//netcode/current/advanced-topics/transports},
  urldate = {2022-05-28},
  langid = {english}
}

@online{UnderstandingHowFacebook2021,
  title = {Understanding {{How Facebook Disappeared}} from the {{Internet}}},
  date = {2021-10-04T21:08:52},
  url = {http://blog.cloudflare.com/october-2021-facebook-outage/},
  urldate = {2022-05-20},
  langid = {english},
  organization = {{The Cloudflare Blog}}
}

@online{UninvitedInternetThings2021,
  title = {The Uninvited {{Internet}} of Things [{{LWN}}.Net]},
  date = {2021},
  url = {https://lwn.net/Articles/850218/},
  urldate = {2022-01-03}
}

@online{UnityTutorial,
  title = {Unity {{Tutorial}}},
  url = {https://learn.unity.com/},
  urldate = {2022-05-31},
  organization = {{Unity Learn}}
}

@online{VirtualAugmentedReality,
  title = {Virtual and {{Augmented Reality}}},
  url = {https://ryan.freumh.org/blog/vr_ar/},
  urldate = {2022-05-15}
}

@article{wardNewLocationTechnique1997,
  title = {A New Location Technique for the Active Office},
  author = {Ward, Andy and Jones, Alan and Hopper, A.},
  date = {1997},
  journaltitle = {IEEE Wirel. Commun.},
  doi = {10.1109/98.626982}
}

@online{WeirdWonderfulWorld,
  title = {The Weird and Wonderful World of {{DNS LOC}} Records},
  url = {https://blog.cloudflare.com/the-weird-and-wonderful-world-of-dns-loc-records/},
  urldate = {2022-06-01}
}

@article{weiserComputer21stCentury1999,
  title = {The Computer for the 21st Century},
  author = {Weiser, Mark},
  date = {1999-07},
  journaltitle = {ACM SIGMOBILE Mobile Computing and Communications Review},
  shortjournal = {SIGMOBILE Mob. Comput. Commun. Rev.},
  volume = {3},
  number = {3},
  pages = {3--11},
  issn = {1559-1662, 1931-1222},
  doi = {10.1145/329124.329126},
  url = {https://dl.acm.org/doi/10.1145/329124.329126},
  urldate = {2021-12-27},
  langid = {english}
}

@article{weiserComputerScienceIssues1993,
  title = {Some Computer Science Issues in Ubiquitous Computing},
  author = {Weiser, Mark},
  date = {1993-07},
  journaltitle = {Communications of the ACM},
  shortjournal = {Commun. ACM},
  volume = {36},
  number = {7},
  pages = {75--84},
  issn = {0001-0782, 1557-7317},
  doi = {10.1145/159544.159617},
  url = {https://dl.acm.org/doi/10.1145/159544.159617},
  urldate = {2022-05-13},
  langid = {english}
}

@article{weiserDesigningCalmTechnology1996,
  title = {Designing {{Calm Technology}}},
  author = {Weiser, Mark and Brown, John Seely},
  date = {1996},
  journaltitle = {Powergrid Journal},
  volume = {1}
}

@software{WelcomeWindowsTerminal2022,
  title = {Welcome to the {{Windows Terminal}}, {{Console}} and {{Command-Line}} Repo},
  date = {2022-05-28T15:01:36Z},
  origdate = {2017-08-11T18:38:22Z},
  url = {https://github.com/microsoft/terminal},
  urldate = {2022-05-28},
  organization = {{Microsoft}}
}

@online{WindowsTerminalArchitecture,
  title = {Windows {{Terminal}} Architecture Doesn't Work in {{Mixed Reality}} Environment · {{Issue}} \#693 · Microsoft/Terminal},
  url = {https://github.com/microsoft/terminal/issues/693},
  urldate = {2022-05-28},
  langid = {english},
  organization = {{GitHub}}
}

@unpublished{xiaEthereumNameService2021,
  title = {Ethereum {{Name Service}}: The {{Good}}, the {{Bad}}, and the {{Ugly}}},
  shorttitle = {Ethereum {{Name Service}}},
  author = {Xia, Pengcheng and Wang, Haoyu and Yu, Zhou and Liu, Xinyu and Luo, Xiapu and Xu, Guoai},
  date = {2021-04-11},
  number = {arXiv:2104.05185},
  eprint = {2104.05185},
  eprinttype = {arxiv},
  primaryclass = {cs},
  publisher = {{arXiv}},
  doi = {10.48550/arXiv.2104.05185},
  url = {http://arxiv.org/abs/2104.05185},
  urldate = {2022-06-05},
  archiveprefix = {arXiv}
}

\end{document}